\documentstyle[preprint,eqsecnum,aps,epsf]{revtex}
\newif\iftightenlines\tightenlinesfalse
\tightenlines\tightenlinestrue
\def\mz{M_Z}
\def\mgut{M_{GUT}}
\def\del{\delta}
\def\eslt{\not\!\!{E_T}}
\def\to{\rightarrow}

\def\te{\tilde e}

\def\tb{\tilde b}

\def\tst{\tilde t}

\def\tg{\tilde g}

\def\tq{\tilde q}
\def\tw{\widetilde W}
\def\tz{\widetilde Z}
\begin{document}
\draft
\preprint{\vbox{\baselineskip=14pt%
   \rightline{FSU-HEP-981015}\break 
   \rightline{UH-511-925-99}\break
   \rightline{NU-TH-99-50}
}}
\title{
THE REACH OF FERMILAB TEVATRON UPGRADES \\
FOR SU(5) SUPERGRAVITY MODELS \\
WITH NON-UNIVERSAL GAUGINO MASSES}
\author{Greg Anderson$^1$, Howard Baer$^2$, Chih-Hao Chen$^2$
and Xerxes Tata$^{3}$}
\address{
$^1$Department of Physics,
Northwestern University,
Evanston IL 60208 USA
}
\address{
$^2$Department of Physics,
Florida State University,
Tallahassee, FL 32306 USA
}
\address{
$^3$Department of Physics and Astronomy,
University of Hawaii,
Honolulu, HI 96822 USA
}
\date{\today}
\maketitle
\begin{abstract}

We explore the reach of luminosity upgrades of 
the Fermilab Tevatron collider for
$SU(5)$ supergravity models in which non-universal GUT-scale gaugino masses
arise via a vacuum expectation value for the auxiliary component of a
superfield that transforms as a {\bf 24, 75 or 200} dimensional
representation of $SU(5)$. This results in a
different pattern of sparticle masses and mixing angles from what is 
expected in
the minimal supergravity model (mSUGRA) with universal $GUT$ scale
gaugino masses. We find that the resulting signal cross sections, and
hence the reach of the Tevatron, are sensitive to the gaugino masses at
the GUT scale. 
In the {\bf 24} model, the large splitting amongst the two lightest neutralinos
leads to SUSY events containing many isolated leptons, including 
events with a real leptonic $Z$ boson plus jets plus
missing energy signal which is visible over much of parameter space. 
In contrast, in the {\bf 75} and {\bf 200}
models, the reach via leptonic SUSY signals is greatly
reduced relative to mSUGRA, and
the signal is usually visible only via the canonical $\eslt +$jets channel. 

\end{abstract}

\medskip



\section{Introduction}

The minimal supergravity model (mSUGRA)~\cite{sugra} provides a
well-motivated and economical framework in which to embed the Minimal
Supersymmetric Standard Model\cite{mssm}, or MSSM.  
In mSUGRA,
supersymmetry is broken in a hidden sector, and 
SUSY breaking is communicated to the visible
sector MSSM fields via interactions of gravitational strength. 
Motivated by
the apparently successful gauge coupling unification in the MSSM, it is
usually assumed that this leads to a common value $m_0$ for all scalars, a 
common mass $m_{1/2}$ for all gauginos, and a common trilinear SUSY 
breaking term $A_0$ at the scale 
$M_{GUT}\simeq 2\times 10^{16}$ GeV. 
The soft SUSY breaking terms, the gauge and Yukawa couplings
and the supersymmetric $\mu$ term are all then evolved from $M_{GUT}$ to
some scale $M\simeq M_{weak}$ using renormalization group equations
(RGE's). 
Electroweak symmetry is broken radiatively due to the large top quark 
Yukawa coupling.
The resulting weak scale spectrum of
superpartners and their couplings can then be derived in terms of four
continuous plus one discrete parameters
\begin{equation}
m_0,\ m_{1/2},\ A_0,\ \tan\beta\ {\rm and} \mathop{\rm sgn}(\mu),
\end{equation}
in addition to the usual parameters of the standard model.

In studies of mSUGRA and other supersymmetric extensions of the
standard model based on gauge-unification and the gravitational
mediation of supersymmetry-breaking, it is often assumed 
(as discussed above) that the
unification of gauge interactions implies a similar unification of
gaugino masses at the scale of gauge-coupling unification.  However,
gravitationally mediated supersymmetry breaking may lead to
non-universal gaugino masses even in the presence of gauge coupling
unification.  We present a class of models which contain non-universal
gaugino masses, discuss their experimental signatures at the Fermilab
Tevatron, and contrast those signatures with those of mSUGRA.  The
models we discuss represent equally predictive alternatives to the
canonical universal gaugino mass scenario.

If gravity is the messenger which communicates supersymmetry breaking 
from the hidden to the visible sector, supersymmetry breaking mass terms 
for gauginos can arise from higher dimensional interactions  
which couple a chiral superfield to the supersymmetric 
field strength\cite{Cremmer}.  
These interactions arise from the locally supersymmetric gauge 
field strength interactions:
\begin{equation}
{\cal L} \supset \int d^4 \theta E(R^{-1}f_{ab}W^a W^b + h.c)
\end{equation}
with a 
gauge kinetic function 
$f_{AB} = \delta_{AB} + \Phi_{AB}/M_{Planck} + \ldots$. 
The fields
$\Phi_{AB}$ transform as left handed chiral superfields
under supersymmetry transformations, and as
the symmetric product of two adjoints under gauge symmetries.
The lowest order contribution to gaugino masses arising from the interaction
above comes from a dimension five operator:

\begin{equation}
{\cal L}\supset \int d^2\theta {W}^a{W}^b 
{\Phi_{ab}\over M_{\rm Planck}} + h.c.
\supset  {\langle F_{\Phi} \rangle_{ab}\over M_{\rm Planck}}
\lambda^a\lambda^b\, +\ldots ,
\end{equation}
where the $\lambda^{a,b}$ are the gaugino fields, 
and $F_{\Phi}$ is the auxillary field component of ${\Phi}$.


In conventional models of supersymmetry breaking,
the fields $F_{\Phi}$ which break supersymmetry are treated as gauge singlets.
However, in principle, the chiral superfield which communicates 
supersymmetry breaking to the gaugino fields can lie in any
representation found in the symmetric product
of the adjoint.
Non gauge singlet vacuum expectation values for the supersymmetry preserving 
component of $\Phi_{AB}$ have been considered previously
\cite{Hill,SW} for their perturbative effect on gauge coupling unification
and also for their effect on gaugino masses\cite{Hill,ellis,drees,anderson}.
Here we consider the effect of supersymmetry breaking vacuum
expectation values of $\Phi_{AB}$ which lead to maximally predictive 
gaugino masses.
In the context of $SU(5)$ grand
unification, 
$F_{\Phi}$
belongs to an $SU(5)$ irreducible representation which
appears in the symmetric product of two adjoints:
\begin{equation}
({\bf 24}{\bf \times}
 {\bf 24})_{\rm symmetric}={\bf 1}\oplus {\bf 24} \oplus {\bf 75}
 \oplus {\bf 200}\,,
\label{irrreps}
\end{equation}
where only $\bf 1$ yields universal masses.  Only the component of
$F_{\Phi}$ that is `neutral' with respect to the SM gauge group should
acquire a vaccuum expectation value (vev), $\langle F_{\Phi}
\rangle_{ab}=c_a\del_{ab}$, with $c_a$ then determining the relative
magnitude of the gaugino masses at $\mgut$.  The relations amongst the
various GUT scale gaugino masses have been worked out, {\it e.g.} in
Ref. \cite{anderson}.  The relative $GUT$ scale $SU(3)$, $SU(2)$ and
$U(1)$ gaugino masses $M_3$, $M_2$ and $M_1$ are listed in
Table~\ref{masses} along with the approximate masses after RGE evolution
to $Q\sim M_Z$.  Here, motivated by the measured values of the
gauge couplings at LEP, we assume that the vev of the scalar component
of ${\Phi}$ is neglible. In principle, as shown in
Fig.~\ref{fig0}, an arbitrary linear combination of the above
irreducible representations is also allowed. We 
consider the implications of models where the dominant contribution
to gaugino masses arises from a single irreducible representation.~\footnote{From the point of view of the theory below the GUT
scale, we may consider the use of the large representations listed in
Eq.~(1.4) as a calculational convenience.  Only the $SU(3)\times
SU(2)\times U(1)$ singlet components of these representations are
relevant to our discussion.  The remaining states may obtain masses
which are heavy compared to $M_{GUT}$.  Any relic of a large GUT
representation which survives below the GUT scale and has a
non-vanishing coupling to the supersymmetric field strength, must lie in
the symmetric product the adjoint representations of the unified group,
and further this relic must be a $SU(3)\times SU(2) \times U(1)$
singlet.  For the unification group $SU(5)$, the complete set of masses
produced by relics from pure $SU(5)$ representations are those listed in
Table 1.  } Each of the three non-singlet models is as predictive as the
canonical singlet case, and all are compatible with the unification of
gauge couplings.  These scenarios represent the predictive subset
of the more general case\cite{drees} of an arbitrary superposition 
of these
representations\footnote{ In
Ref.\cite{ellis}, a specific linear combination was fixed by the
additional assumption of the vanishing of this contribution to
leptoquark gaugino masses.}, the most interesting being a superposition
of gauge singlet and adjoint fields. 

As we discuss in Section III, signals of supersymmetry-- and hence 
the reach of the Tevatron-- is sensitive to the structure of the gaugino
masses as the GUT scale.
The reach of the Fermilab Tevatron collider for mSUGRA models with
universal gaugino masses has been worked out for both
low\cite{bcpt,mrenna,tev33,snow} as well as high\cite{bcdpt} values of
the parameter $\tan\beta$.  For low values of $\tan\beta$ and high
integrated luminosity, the clean trilepton signal (C3L)\cite{trilep}
from $\tw_1\tz_2\to \ell'\bar{\ell}\ell+\eslt$ usually offers the best
prospect for a SUSY discovery. For parts of parameter space, a SUSY
signal might be found in several different channels.  For large
$\tan\beta$, discovery via the C3L signal becomes increasingly difficult
because sparticle decays to $\tau$-leptons and $b$-quarks becomes
enhanced relative to decays to $e$'s and $\mu$'s, but the range of
parameters over which the signal is observable may be
extended by the use of softer cuts on the leptons as 
emphasized by Barger {\it et al.}\cite{bkl}.

We should add that we do not specially advocate any
particular representation for $\langle F_{\Phi} \rangle$ on theoretical
grounds. Our main motivation is to examine the sensitivity of the various
signals via which SUSY might manifest itself at future runs of the
Tevatron to changes in the underlying framework. It is especially
important to do so when assessing the search capabilities of future
facilities, particularly because we do not as yet have a
dynamical understanding of SUSY breaking, which can affect the
phenomenology via the pattern of sparticle masses and
mixing angles.

With this in mind, the event generator ISAJET\cite{isajet} 
(versions $\ge 7.37$)
has been upgraded to accommodate SUGRA models with various non-universal
soft SUSY breaking terms. In this study, we use ISAJET to simulate models with
non-universal gaugino mass parameters at the scale $M_X$ assuming
universality of other parameters.
The model parameter space used in this paper thus
corresponds to
\begin{equation}
m_0,\ M_3^0,\ A_0,\ \tan\beta\ {\rm and} \mathop{\rm sgn}(\mu),
\end{equation}
where $M_i^0$ is the $SU(i)$ gaugino mass at scale $Q=M_{GUT}$.  $M_2^0$
and $M_1^0$ can then be calculated in terms of $M_3^0$ according to
Table \ref{masses}. ISAJET calculates an iterative solution to the 26
RGEs, and imposes the radiative electroweak symmetry breaking
constraint. This determines all sparticle masses and mixings. Next,
branching fractions for all sparticles, particles and Higgs bosons are
calculated. Supersymmetric particle production events can be generated
for all possible $2\to 2$ SUSY hard scattering subprocesses. Sparticle
production is followed by initial and final state parton showers,
cascade decays, hadronization and underlying event simulation. Thus,
specific assumptions about soft SUSY breaking terms that are motivated
by $GUT$ or $String$ scale physics can be directly tested at
collider experiments.

In this paper, we explore the consequences of non-universal gaugino masses
for $SU(5)$ SUGRA GUT models for the Fermilab Tevatron collider and its 
planned upgrades. Our goals are several.
\begin{itemize}
\item We wish to establish the capability of the Tevatron and its upgrades
to discover or rule out SUSY within the context of models alternative to
mSUGRA. The set of models we examine maintain many of the attractive 
features of generic SUGRA models, while exhibiting radically different 
sparticle mass spectra and mixing angles
from the commonly examined models which assume
universality.
\item We want to see if this class of models examined can be distinguished 
one from another. If certain SUSY signals are observed, the answer appears to
to be yes for a limited region of model parameter space.
\item Are there any new signals for SUSY that can occur within the context
of non-standard SUSY models? We will see that in the $F_\Phi\sim {\bf 24}$ 
model, there is a large range of parameter space that leads to signal events
containing real leptonic $Z$ bosons. These signals occur much more rarely 
in the mSUGRA model.
\end{itemize}
Non-universality of gaugino masses can also arise in other model 
contexts\cite{moroi} including some string models\cite{ibanez}.
Phenomenological consequences of O-II string models have been
examined in Ref. \cite{cdg}.

In Sec. II, we outline features of the mass spectra that are 
consequences of the assumptions about the $SU(5)$ 
representation of the hidden sector field(s) $\Phi$ that can occur.
In Sec. III, we outline the various types of signals that could occur for
SUSY models, and our signal and background event generator calculations.
In Sec. IV, we present results of the reach of Tevatron upgrade options 
for each of the four models considered. In Sec. V, we present a summary 
and some conclusions.

\section{Sparticle masses for SUGRA models}

We begin by illustrating the evolution of the magnitude of 
soft SUSY breaking masses versus scale $Q$
in Fig. \ref{FIG1} for the four model choices {\it a}) $F_\Phi\sim{\bf 1}$, 
{\it b}) $F_\Phi\sim {\bf 24}$, {\it c}) $F_\Phi\sim {\bf 75}$ and
{\it d}) $F_\Phi\sim {\bf 200}$. 
We take $m_0=100$ GeV, $M_3^0=125$ GeV, $A_0=0$, $\tan\beta =5$ and
$\mu >0$. Throughout this paper, we take $m_t=175$ GeV.

The gaugino masses are denoted by dashed lines, while Higgs masses are
denoted by dotted lines and squark and slepton masses are denoted by
solid lines. For the usual mSUGRA case illustrated in
Fig. \ref{FIG1}{\it a}, the gaugino masses evolve from a common GUT
scale value. For the $F_\Phi \sim {\bf 24}$ model in frame {\it b}), the
splitting in GUT scale gaugino masses shown in Table I
leads to a large mass gap between
$M_1$ and $M_2$ at the weak scale, and also a large mass gap between
left and right sfermions. In case {\it c}) for $F_\Phi \sim
{\bf 75}$, the large GUT scale splitting of gaugino masses leads to near
gaugino mass degeneracy at the weak scale, and also similar masses for 
both squarks and sleptons. Finally, for case {\it d}) with 
$F_\Phi\sim {\bf 200}$, the large GUT scale splitting leads to 
$M_2,M_3 <M_1$ at the weak scale.
In addition, the large GUT scale values of $M_1$ and $M_2$ cause the weak
scale slepton masses to evolve to relatively high masses compared to the
$F_\Phi \sim {\bf 1}$ and ${\bf 24}$ models, so that
left sfermions are lighter than
right sfermions; this is in contrast to usual expectations from models
with universal gaugino masses. 
The $m_{H_2}^2$ mass
parameter initially has an upward trajectory, but is ultimately evolved
to negative values so that radiative electroweak symmetry is just barely
broken. 

A variety of physical sparticle masses along with the magnitude of the
weak scale $\mu$ parameter are shown versus $\tan\beta$ in
Fig. \ref{FIG2} for the four model choices using the same parameters as
in Fig. \ref{FIG1}.  Frame {\it a}) shows the generic mSUGRA model
spectrum for comparison with the models with non-universal gaugino
masses. 
In frame {\it b}), the large mass gap between $m_{\tw_1}$ or
$m_{\tz_2}$ and $m_{\tz_1}$ is apparent.  This mass gap has important
consequences for collider experiments: frequently it is so large that
neutralino decays to real $Z$ bosons are often allowed! Signatures
involving real $Z$s could be a distinctive signature for models leading
to large mass gaps between $\tz_2$ and $\tz_1$.  

For the $F_\Phi \sim
{\bf 75}$ case in frame {\it c}), there is almost no mass gap between
$m_{\tw_1}$ and $m_{\tz_1}$. For instance, for $\tan\beta=5$, with
the other parameters as in the figure, $m_{\tw_1}-m_{\tz_1}$ is just
0.5~GeV; this gap increases slightly with $\tan\beta$. The mass
difference between $\tz_2$ and $\tz_1$ though larger ($\sim 18$~GeV for
$\tan\beta=5$) is still considerably smaller than in the canonical mSUGRA case.
In this case, decays of $\tz_2$ and certainly $\tw_1$ will lead to very
soft visible particles which will make detection of hard isolated
leptons from cascade decays very difficult. In view of the very tiny
mass difference between the chargino and $\tz_1$, the reader may
legitimately wonder whether the chargino is sufficently long lived as to
travel a substantial distance in the detector, thus leaving a track
before decaying. We have checked, however,~\footnote{Because the
mass gap is smaller than 1~GeV, it is not reasonable to compute the
hadronic decay width of the chargino using
ISAJET, which really computes the decay $\tw_1 \to
q\bar{q}\tz_1$. Instead it is more reasonable to compute exclusive
decays into 1,2, {\it etc.} pion states in association with $\tz_1$. We
are grateful to M.~Drees who has provided us a code to do so. For the
$\tan\beta =5$ point discussed in the text, the lifetime using this code
agrees with the ISAJET lifetime to within a factor 2. For the decay of
$\tz_2$ for which the mass gap is $\sim 20$~GeV, the decays can, of
course, be calculated using ISAJET.}
that the lifetime of the
chargino is ${\cal O}(10^{-11}$~s) so that this appears not to be the
case. 

For the $F_\Phi \sim {\bf 200}$ model in frame {\it d}), the
$\tw_1$-$\tz_1$ mass gap is just a few GeV, while
$m_{\tz_2}-m_{\tz_1}$ is several tens of GeV. We have checked, however,
that $\tau_{\tw_1} \agt 10^{-15}$ seconds, so that it decays rapidly
without an appreciably displaced vertex. In this case, the $\tz_4$ is
mainly a bino, and is the heaviest of all the sparticles.

Aside from that alteration of the masses, the weak scale values of the
gaugino masses in Table \ref{masses} also imply very different mixing
patterns for the charginos and neutralinos as compared to the usual
mSUGRA case. In contrast to the mSUGRA case, $|\mu|$ tends to be
somewhat smaller than $M_2$, and 
the lighter neutralinos and $\tw_1$ are
dominantly Higgsino-like in the {\bf 75} and {\bf 200} cases. This 
impacts on the decays of sparticles, {\it e.g.} $\tz_2$ and 
sometimes also $\tw_1$ production in cascade decays tends to be suppressed,
while frequently heavier charginos and neutralinos
are produced with large rates. The decay patterns of $\tw_1$ and $\tz_2$
are also changed from usual mSUGRA expectation. This will reflect itself
in changes in expected rates for various event toplogies as we will see later.

The different boundary condition for gaugino masses sometimes has a
strong effect on other masses via the RGE.
For instance, for the $F_\Phi \sim {\bf 200}$ model shown in frame {\it d}),
the huge GUT scale value of $M_1=1250$ GeV causes right slepton and
squark masses to evolve to large values so that in this case
$m_{\te_R}>m_{\tq}>m_{\te_L}$!
Another significant difference from the usual mSUGRA case is the large
splittings between the masses of various squarks in the {\bf 75} and
{\bf 200} cases. Indeed it is sometimes possible to have $m_{\tq_L} \geq
m_{\tg} \geq m_{\tq_R}$, so that gluinos decay almost exclusively to
right handed squarks. This, in turn, alters the cascade decay patterns
from usual expectation because the right handed squarks cannot decay
into charginos and neutralinos with dominant $SU(2)$ components.


In Fig. \ref{FIG3}, we show gluino and squark mass contours
in the $m_0\ vs.\ M_3^0$ plane for $\tan\beta =5$, $A_0=0$ and $\mu >0$.
The bricked regions are excluded by theoretical constraints:
either electroweak symmetry is not broken appropriately, or the
lightest SUSY particle (LSP) is not the lightest neutralino, in 
contradiction with results from searches for stable cosmological
relics. These regions are sensitive to the exact choice of $m_t$. 
The gray shaded regions are excluded by collider search 
experiments for SUSY particles, and are mainly formed from the LEP2
bound that $m_{\tw_1} >85.5$ GeV\cite{lep2}; the LEP2
bound from the non-observation of $h$ plays a smaller role since
for $\tan\beta =5$, $m_h$ is usually not small. The chargino bounds used may 
actually be
too stringent for the $F_{\Phi}\sim {\bf 75}$ and {\bf 200} models where
the $m_{\tw_1}-m_{\tz_1}$ mass gap is small; for these cases, the LEP2
limits will have to be re-analyzed.~\footnote{Since the two lighter
neutralinos contain significant Higgsino components, and
$m_{\tz_2}-m_{\tz_1}$ is at least a few GeV for the {\bf 75} model
(tens of GeV for the {\bf
200} model), we may expect LEP experiments might be able to detect
signals from $e^+e^- \to \tz_1\tz_2$ production. For the {\bf 200} case,
the non-observation of acollinear leptons or jets from $\tz_2$ decay
could lead to significant limits on its mass. In the {\bf 75} case the
analysis will have to be redone since $m_{\tz_2}-m_{\tz_1}$ is just a
few GeV, but it is worth keeping in mind that in the MSSM, ALEPH finds
a mass bound of 79~GeV
on $m_{\tw_1}$ that is derived by combining chargino and neutralino
searches, assuming a mass gap $\geq 5$~GeV. Finally, we note that in
the {\bf 75} scenario, the branching fraction for the decay $\tz_2 \to
\tw_1 \ell\nu$ is significant; since the daughters of $\tw_1$ are likely
to be soft, $\tz_1\tz_2$ production could result in ``monolepton'' events at
LEP. While it is clear that a dedicated analysis is required to really
exclude the ``hatched region''  for the {\bf 200}, and especially the
{\bf 75} cases, we have chosen to show it using the same criteria in all
four cases.} 
The gluino and squark mass contours are intended for comparison with
the parameter space reach plots that will be presented in
Sec. 3 of this paper. In Fig. \ref{FIG4}, we show the same mass 
contours for $\tan\beta =25$. In this case, the parameter space is much more 
restrictive. In particular, for the $F_{\Phi}\sim {\bf 75}$ 
model in frame {\it c}),
radiative electroweak symmetry breaking is difficult to achieve for 
large values of the parameter $M_3^0$.

\section{Event simulation and reach calculations}

In several previous studies\cite{bcpt,mrenna,tev33,snow,bcdpt}, a
variety of signal channels for the discovery of mSUGRA (with universal
GUT scale gaugino masses) at the Tevatron were investigated, and the
reach of the Tevatron Main Injector era (MI-integrated luminosity of 2
fb$^{-1}$) and TeV33 era (integrated luminosity of $\sim 25 $fb$^{-1}$)
were delineated in the parameter space of the mSUGRA model. We had
investigated\cite{bcpt,bcdpt} several promising discovery channels that
included
\begin{itemize}
\item multi-jet $+\eslt$ events (veto hard, isolated leptons) (J0L),
\item events with a single isolated lepton plus jets $+\eslt$ (J1L),
\item events with two opposite sign isolated leptons plus jets $+\eslt$ (JOS),
\item events with two same sign isolated leptons plus jets $+\eslt$ (JSS),
\item events with three isolated leptons plus jets $+\eslt$ (J3L),
\item events with two isolated leptons $+\eslt$ (no jets, clean) (COS),
\item events with three isolated leptons $+\eslt$ (no jets, clean) (C3L).
\end{itemize}
In these samples, the number of leptons is {\it exactly} that indicated,
so that these samples are non-overlapping.
For Tevatron data samples on the order of 0.1 fb$^{-1}$, the J0L
signal generally gave the best reach for supersymmetry. It is the classic 
signature for detecting gluinos and squarks at hadron colliders. For 
larger data samples typical of those expected at the MI or TeV33, 
the C3L signal usually yielded the greatest reach except when leptonic
decays of charginos and neutralinos are strongly suppressed. 
In the present paper, we will
extend these results for models with non-universal $GUT$ scale gaugino 
masses.

We have found that the second model described above 
with $F_\Phi \sim{\bf 24}$ can give rise to
SUSY events at the Tevatron which are rich in $Z$ bosons. To extract
this signal, we require:
\begin{itemize}
\item identification of a leptonic ``$Z$'' boson ($Z\to e^+e^-$ or
$\mu^+\mu^-$) plus jets plus $\eslt$ (JZ).
\end{itemize}

To model the experimental conditions at the Tevatron, we use the toy
calorimeter simulation package ISAPLT. We simulate calorimetry covering
$-4<\eta <4$ with cell size $\Delta\eta\times\Delta\phi =0.1\times
0.0872$. We take the hadronic (electromagnetic) energy resolution to be
$70\% /\sqrt{E}$ ($15\% /\sqrt{E}$).  Jets are defined as hadronic
clusters with $E_T > 15$~GeV within a cone with $\Delta
R=\sqrt{\Delta\eta^2 +\Delta\phi^2} =0.7$. We require that $|\eta_j|
\leq 3.5$.  Muons and electrons are classified as isolated if they have
$p_T>5$~GeV, $|\eta (\ell )|<2.5$, and the visible activity within a
cone of $R=0.3$ about the lepton direction is less than $max(E_T(\ell
)/4,2\ {\rm GeV})$.  
In our analysis, we neglect multiple scattering effects,
non-physics backgrounds from photon or jet misidentification, and make
no attempt to explicitly simulate any particular detector.

We incorporate in our analysis the following trigger conditions:
\begin{enumerate}
\item one isolated lepton with $p_T(\ell) > 15$~GeV and $\eslt >15$ GeV,
\item $\eslt >35$ GeV,
\item two isolated leptons each with $E_T>10$ GeV and $\eslt >10$ GeV,
\item one isolated lepton with $E_T>10$ GeV plus at least one jet plus
$\eslt >15$ GeV,
\item at least four jets per event, each with $E_T>15$ GeV.
\end{enumerate}
Thus, every signal or background event must satisfy at least one of the 
above conditions. 
In addition to these basic selection and trigger criteria, we impose
various additional cuts listed in Ref. \cite{bcpt}
the various signal classes. In particular, 
for the jetty channels, we require $E_T(j_1), E_T(j_2)$ and $\eslt$ all
to exceed a cut parameter $E_T^c$ which is chosen to maximize the reach,
while for the clean trilepton (C3L)
channel, we require rather hard leptons with $E_T(\ell_1,\ell_2,\ell_3)
\geq (20, 15, 10)~$~GeV.

We have generated the following physics background processes using ISAJET:
$t\bar t$ production, $W+$jets, $Z+$jets, $WW$, $WZ$ and $ZZ$ production
and QCD (mainly from $b\bar b$ and $c\bar c$ production). Each 
background subprocess was generated with the hard scattering
subprocess final state particles
in $p_T$ bins of $25-50$ GeV, $50-100$ GeV, $100-200$ GeV, $200-400$ GeV
and $400-600$ GeV. 
 The numerical background values we use are listed in 
Ref. \cite{bcpt}, and will not be repeated here.

For the new JZ event channel, we require two opposite sign same flavor
isolated leptons ($e$ or $\mu$) with $m(\ell\bar{\ell})$ within $M_Z\pm 8$GeV.
We also require $n(jets)\ge 2$, $S_T\ge 0.2$ and $\eslt \ge 40$ GeV.
In this case, the background rate was found to be 13.6 fb, mostly coming
from $t\bar{t}$, $WZ$ and $ZZ$ production.

\section{Tevatron reach results for SUGRA models with non-universal 
gaugino masses}

In Fig. \ref{FIG5} -- Fig. \ref{FIG20}, we show the results of our
computation of the SUSY reach of  
Tevatron collider experiments for models with non-universal gaugino
masses.  For each set of model input parameters, and for a given integrated
luminosity, we consider a signal to be observable above background if,
(for some value of the cut parameter $E_T^c$ for the jetty channels
other than the $JZ$ channel)
\begin{itemize}
\item $S>5\sqrt{B}$,
\item $S>0.2 B$, and
\item $S>5$ (10) for integrated luminosity equal to
$0.1$ or $2$ fb$^{-1}$ (25 fb$^{-1}$),
\end{itemize} 
where $S$ is the expected number of signal events and $B$ is the
expected number of background events. Within our framework, the scale
of sparticle masses (and hence their production rates) is mainly
determined by the parameters $m_0$ and $M_3^0$ (which fixes other
gaugino masses at the unification scale). For this reason, the
$m_0-M_3^0$ plane provides a convenient way to present our results. 
The results are somewhat less sensitive to variation of other
parameters. In our analysis, we fix $A_0=0$ and choose $\mu > 0$
(negative values of $\mu$ frequently do not yield the correct symmetry
breaking pattern), and illustrate our results for $\tan\beta=5$ and 25.
Sampled points for which there is an observable signal 
for integrated luminosity of 0.1
fb$^{-1}$ are denoted by black squares; gray squares
denotes points where the signal is observable with 2 fb$^{-1}$ 
and white squares, points that can be accessed with 25
fb$^{-1}$. Sampled points not accessible with even 25~fb$^{-1}$
of  integrated luminosity are denoted with an $\times$.

\subsection{Reach via the J0L channel}

Fig. \ref{FIG5} shows results in the $m_0\ vs.\ M_3^0$ plane for
$\tan\beta =5$, $A_0=0$ and $\mu >0$.  These results are in the J0L
channel, which is the classic signature for supersymmetry at hadron
colliders.  For the $F_{\Phi}\sim {\bf 1}$ case with universal gaugino
masses in frame {\it a}), we find no reach for mSUGRA (black squares)
with the current Tevatron data sample beyond the region already excluded
at LEP2.  However, experiments at the MI should be able to probe
$M_3^0$ values up to 150~GeV ($m_{\tg}\simeq 400$ GeV) for lower values of
$m_0$. The TeV33 integrated luminosity extends this reach to
$M_3^0=175$ GeV, corresponding to $m_{\tg}\simeq 480$ GeV. For the
$F_{\Phi}\sim{\bf 24}$ model in frame {\it b}), there is a significant
reach of {\it current} Tevatron experiments beyond the reach of
LEP2. This is due mainly to the increased values of $m_{\tw_1}$ and
$m_{\tz_2}$ relative to their mSUGRA counterparts for a given value of
$M_3^0$, so that just beyond the LEP2 limit, relatively light values of
$m_{\tg}\sim 300$ GeV are still allowed, and can give rise to large J0L
signals. The overall reach for SUSY in frame {\it b}) extends to
$M_3^0=175$ GeV, which is comparable to the mSUGRA case in frame {\it
a}). For these large values of $M_3^0$, gluinos and squarks are heavy,
and chargino and neutralino production is the dominant SUSY mechanism. 
For $M_3^0=175$~GeV, $m_{\tw_1}$ is significantly heavier in the {\bf 24}
model relative to the mSUGRA case: the accessibility of heavier
charginos is presumably due to the larger mass gap between the chargino
and the LSP, which should increase the efficiency for detecting J0L events.
For the $F_{\Phi}\sim{\bf 75}$ model in frame {\it c}), the limits
from LEP2 are again suppressed compared to the mSUGRA case due to
heavier values of $m_{\tw_1}$ for a given value of $M_3^0$. In this
model, $m_{\tw_1}\simeq m_{\tz_1}$ so that there is very little visible
energy from $\tw_1$ decays, and they behave effectively like the $\tz_1$
in the detector, {\it i.e.} they give missing energy.  Gluino and squark
pair production gives rise to a significant J0L signal for low values of
$M_3^0$, so that there is still a substantial reach for SUSY via the MI
and TeV33. The reach of TeV33 is somewhat smaller than in the mSUGRA
case because for values of $M_3^0\simeq 175$ GeV, direct $\tw_1$ and
$\tz_2$ production dominates $\tg\tg$ and $\tg\tq$ production: {\it
e.g.} for mSUGRA, $\tw_1\overline{\tw_1}$ production leads to
jets$+\eslt$ events, but for the $F_{\Phi}\sim{\bf 75}$ model no hard
jets get produced in $\tw_1$ decay. Finally, the reach for the
$F_{\Phi}\sim{\bf 200}$
model is shown in frame {\it d}).  In this model, as in the {\bf 75}
case, relatively light values of $m_{\tg}$ are accessible to Tevatron
experiments, and there is a significant reach for SUSY via the J0L
signal. The black squares in the lower left of the frame come mainly
from $\tst_1\bar{\tst_1}$ and $\tg\tg$ events where $\tg\to b\tb_1$, so
that the events are rich in $b$-jets. The ultimate reach of TeV33 again
extends to $M_3^0 =150$ GeV for low $m_0$, for which $m_{\tg}\simeq 400$
GeV.

Similar results for the reach of the Tevatron via the J0L channel are
shown in Fig. \ref{FIG6} for $\tan\beta =25$ (all other parameters are
the same). The reach is somewhat diminished from the lower $\tan\beta$
case for all four models.  Nevertheless, we see that there is
significant reach via the Tevatron upgrades for supersymmetry in all
models via the classic J0L channel.

\subsection{Reach via the J1L channel}

In Fig. \ref{FIG7} we show the Tevatron reach via the J1L signal for
$\tan\beta= 5$.  For
the mSUGRA case in frame {\it a}), there is no reach via the MI beyond
the bounds from LEP2, but the TeV33 upgrade can access $M_3^0$ values as
high as $\sim 175$ GeV for some parameter space points.  For the
$F_{\Phi}\sim{\bf 24}$ model in frame {\it b}), the Tevatron MI has
considerable reach for SUSY via the J1L channel beyond the LEP2 bounds.
Much of the reach at lower $M_3^0$ values comes from gluino and squark
cascade decays to $\tw_1$ which then decays leptonically. The large mass
gap between $\tw_1$ and $\tz_1$ (shown in Fig. \ref{FIG2}) results in a
very energetic lepton which has a high probably for detection.  TeV33
can access points with $M_3^0\simeq 175-200$ GeV, corresponding to
$m_{\tg}\sim 500$ GeV. When we next examine the reach in the
$F_{\Phi}\sim{\bf 75}$ and $F_{\Phi}\sim{\bf 200}$ models in frames {\it
c}) and {\it d}), we see no reach via the MI, and only a marginal reach
via TeV33. Much of the signal presumably comes
from cascade decays to $\tz_3$ for which the branching fraction is
substantial --- the $\tz_3$ can then decay into real vector bosons to give the
leptonic signal. In these cases, the small mass gap between $\tw_1$ and
$\tz_1$ yields low energy leptons with a poor probability to pass cuts
in the J1L channel, and furthermore, cascade decays to these states tend
to be somewhat suppressed.

For the $\tan\beta =25$ case in Fig. \ref{FIG8}, in almost all the models,
the reach via the J1L signal is diminished with respect to the lower
$\tan\beta$ cases. Again, this is generally because at high $\tan\beta$,
decays to $b$'s and $\tau$'s are enhanced relative to decays into $e$'s
and $\mu$'s, making SUSY detection via leptonic modes in general more
difficult. The exception here occurs with the $F_{\Phi}\sim{\bf 200}$
model, where there is some reach for the MI beyond the LEP2 bounds. In
this case, some of the J1L events come from cascade decays involving
$\tz_3$ which can decay via $\tz_3\to\tw_1 W$, and a hard lepton results
from the $W$ decay.

\subsection{Reach via the JOS channel}

The Tevatron reach via the JOS channel is illustrated in Fig. \ref{FIG9}
and \ref{FIG10}. For the mSUGRA case in frame {\it a}) of Fig. \ref{FIG9}, 
there is some
reach by the MI and TeV33 for low values of $m_0$ where sleptons become
relatively light, and charginos and neutralinos can directly decay 
to sleptons and sneutrinos. The isolated dileptons come from a variety
of cascade decay mechanisms involving charginos, neutralinos, sleptons and
sneutrinos. For the $F_{\Phi}\sim{\bf 24}$ model in frame {\it b}), there is
a significant reach by Tevatron experiments beyond the LEP2 bounds 
even with the current data sample, and the reach expands considerably 
for the MI and TeV33. The OS dileptons again come from a variety of cascade 
decay mechanisms which include contributions from heavier charginos and 
neutralinos $\tw_2$ and $\tz_3$. In the $F_{\Phi}\sim{\bf 75}$ and {\bf 200}
models, there is {\it no reach} beyond the LEP2 bounds for any Tevatron
luminosity upgrade in this channel. This is, perhaps, not surprising if
indeed the decays $\tz_3 \to W\tw_1$ are the main source of J1L
events, since the $\tw_1$ is mostly invisible; {\it i.e.} any JOS event
is doubly suppressed by the branching fraction of the cascade decay of
gluino or squark into a lepton.
We see a similar pattern for the $\tan\beta =25$ case shown in
Fig. \ref{FIG10}, except also that the reach in the mSUGRA and 
$F_{\Phi}\sim{\bf 24}$ models is diminished due to the
enhancement of decays to $\tau$-leptons and $b$-quarks.

\subsection{Reach via the JSS channel}

The reach for SUSY in the JSS channel has been noted as a distinct
signal for cascade decays of the $\tg$ to $\tw_1$, where the Majorana
nature of the gluino gives rise to equal probability for detection of
same-sign and opposite-sign dileptons. In the mSUGRA model in
Fig. \ref{FIG11}{\it a}), there is only a tiny region that can be
probed at the Tevatron
in this channel mainly because the LEP2 bounds force
$m_{\tg}$ and $m_{\tq}$ to such high values that their production cross
section is suppressed relative to direct chargino, neutralino and
slepton pair production. For the $F_{\Phi}\sim{\bf 24}$ model, however,
lighter values of $m_{\tg}$ are allowed beyond the LEP2 exclusion
region, and furthermore, the large $\tw_1$-$\tz_1$ decay gap gives rise
to a relatively high probability to detect $\tg\to\tw_1\to \ell$ cascade
decay leptons. Consequently, we see a significant reach in the JSS
channel in frame {\it b}). Nonetheless, the reach is somewhat smaller
than in the JOS channel, which also receives significant contributions
from leptonic decays of neutralinos. In the $F_{\Phi}\sim{\bf 75}$ and
{\bf 200} models, there appears to be no signal in the JSS channel
beyond the LEP2 region for much the same reasons that we just discussed
for the JOS case.  Broadly similar results hold for the $\tan\beta =25$ case
illustrated in Fig. \ref{FIG12}, where we see the usual reduction in the
region where there is an observable signal in the mSUGRA and in the
$F_{\Phi}\sim{\bf 24}$ models. 

\subsection{Reach via the J3L channel}

In Fig. \ref{FIG13}, we show the reach of Tevatron experiments in the
J3L channel. There is a significant reach by the Tevatron MI and TeV33
for mSUGRA for $m_0\alt 150$ GeV extending all the way to $M_3^0 = 225$
GeV, corresponding to $m_{\tg}\simeq 600$ GeV, as shown in frame {\it
a}).  The mSUGRA J3L signal dominantly comes from direct chargino,
neutralino, slepton and sneutrino production and decays.  In the
$F_{\Phi}\sim{\bf 24}$ model, there is a significant reach in the J3L
channel even for large values of $m_0$ since $m_{\tg}$ can be as light
as $\simeq 230$ GeV just beyond the LEP2 bound. For the largest values
of $M_3^0$ where there is an observable signal in the mSUGRA and {\bf
24} cases, sparticle production is dominated by chargino and neutralino
production, and the signal dominantly comes from $\tw_1\tz_2$ production
with jets coming from QCD radiation. Because we only require leptons to
have $E_T(\ell) \geq 10$~GeV, we expect that the efficiency increases by
a relatively small amount despite the increase in $m_{\tw_1}-m_{\tz_1}$
in going from the mSUGRA to the {\bf 24} case (in contrast to the case
of the J0L signal where the increase in efficiency might be
substantial). As a result the boundary of the TeV33 region occurs for
similar values of $m_{\tw_1} \sim 170-180$~GeV.  For the
$F_{\Phi}\sim{\bf 75}$ and {\bf 200} models, there is again hardly any
reach for SUSY in the J3L channel. For the large $\tan\beta =25$ case
illustrated in Fig. \ref{FIG14}, the reach for all models in the J3L
channel is diminished due to enhanced decays to 3rd generation
particles, but for the {\bf 24} case, there is still a significant
region beyond the current LEP reach that can be probed at Tevatron
upgrades. 

\subsection{Reach via the C3L channel}

The clean trilepton signal which often comes from 
$\tw_1\tz_2\to 3\ell +\eslt$ has frequently been considered the most
promising signal via which to search for SUSY at luminosity upgrades of
the Tevatron. These analyses have mainly been performed within the
mSUGRA model. But even in this case, it has been known for some time
that there are parameter space regions where there is no observable
signal in this channel because chargino and neutralino decays to leptons
may be suppressed. Our computation of
the reach in this C3L channel is shown in 
Fig. \ref{FIG15} for $\tan\beta =5$. For the mSUGRA model, the reach of the MI
extends to $M_3^0 =200$ GeV, while the TeV33 reach extends past 
$M_3^0 =250$ GeV, corresponding to $m_{\tg}\simeq 650$ GeV! The
well known gap in the reach at $m_0\simeq 200-300$ GeV due to
destructive interference in neutralino leptonic decays is clearly visible.
There is substantial reach for SUSY in the $F_{\Phi}\sim{\bf 24}$ model
both at the MI and TeV33. The reach in $M_3^0$ at low $m_0$ is, however,
diminished relative to the mSUGRA model. This is because the large mass
gap between $m_{\tz_2}$ and $m_{\tz_1}$ allows the spoiler decay
modes $\tz_2\to Z\tz_1$ and $\tz_2\to h\tz_1$ to turn on at lower values
of $M_3^0$. It is, however, interesting to see that there is an
observable signal beyond the LEP bounds, for {\it all} values of $m_0$
scanned in the figure.  
For the $F_{\Phi}\sim{\bf 75}$ and {\bf 200} models, there is
no reach for SUSY in the C3L channel, which underscores the model
dependence of the much touted C3L signal. It may be of interest to
examine whether the use of softer cuts~\cite{bkl} on the leptons
affects this conclusion.

In the large $\tan\beta =25$
case of Fig. \ref{FIG16}, the mSUGRA reach is diminished at low $m_0$
due to enhanced decays to $\tau$ leptons. It may be possible to further enhance
this reach by softening the cuts on the leptons\cite{bkl}.
There remains a significant reach
for SUSY in the C3L channel for the $F_{\Phi}\sim{\bf 24}$ model at large 
$\tan\beta$ because the large mass gap between $m_{\tz_2}$ and $m_{\tz_1}$
allows $\tz_2$ decays to real selectrons and smuons to compete with decays to
staus.

We have also checked the reach via the COS channel. While there are
parameter regions where this could provide confirmation of a signal in
other channels, the COS topology does not appear to increase the reach
beyond what is observable via other channels.

\subsection{Reach via the JZ channel}

It is possible to produce real $Z$ bosons in SUSY particle cascade decay
events. Events with an identified $Z$ boson plus $\eslt$ are interesting
because Standard Model backgrounds to these mainly come from vector
boson pair production or $t\bar{t}$ production where the leptons from
the decays of the tops accidently reconstruct the $Z$ mass, and hence,
are small. Prospects for observing just this signal at the Tevatron
collider were examined long ago in the context of the MSSM framework
\cite{btwz}. In this study, the focus was on relatively small values of
$\mu$ and $m_{\tg}$, so that all the charginos and neutralinos were
accessible via the production and subsequent decays of gluinos.

Although JZ events are possible within the mSUGRA framework, they
typically occur at very low rates, at least for sparticle masses
accessible at the Tevatron. To understand this, we first recall that because 
$|\mu| \gg M_1,\ M_2$ as is typical in mSUGRA, the lighter neutralinos and the
lighter chargino are mainly gaugino-like, while the heavier ones are
higgsino-like. But since the $Z$ boson couples only to higgsino pairs or
charged gaugino pairs, it is clear that 
the widths for the decays $\tz_{3,4} \to \tz_{1,2} Z$ or
$\tw_2 \to \tw_1 Z$ are suppressed by gaugino-Higgsino mixing
angles.

Our results for the observability of SUSY events in the JZ channel are
shown in Fig. \ref{FIG17} for $\tan\beta =5$, and in Fig. \ref{FIG18} for
$\tan\beta =25$. 
Indeed we see from Fig. \ref{FIG17}{\it a} that there is {\it no reach}
at either the Tevatron MI or TeV33 for mSUGRA in the JZ channel.
However, for the $F_{\Phi}\sim{\bf 24}$ model, in frame {\it b}) $|\mu
|\sim M_2$, and the $\tz_3$ can be light enough that it can be directly
produced in collider events, while its decay branching fraction to $Z$
is substantial: $\sim 10-50\%$.  Also, the large $\tz_2-\tz_1$ mass gap
allows the decay $\tz_2\to Z\tz_1$ to occur (via the subdominant
higgsino component of $\tz_1$) in much of parameter space.  We see in
Fig. \ref{FIG17}{\it b}) that while this signal might be detectable at
the MI for a limited range of parameters, the reach of TeV33 in this
channel is indeed substantial, covering much of parameter space below
$M_3^0\alt 150$ GeV!  Meanwhile, for the $F_{\Phi}\sim{\bf 75}$ and {\bf
200} models, there is again no reach for SUSY in the JZ channel -- the
branching fractions for cascade decays to heavier neutralinos and
charginos tend to be small in these cases. 

For the $\tan\beta =25$ case in Fig. \ref{FIG18}, there is again no
reach for SUSY in the mSUGRA model or the $F_{\Phi}\sim{\bf 75}$ and
{\bf 200} models. In the $F_{\Phi}\sim{\bf 24}$ model, there is a
significant Tevatron
reach in the JZ channel, but only for TeV33 type integrated
luminosities.  Since
the JZ signal occurs at an observable level only in rather special 
models, the observation of such a signal in tandem with more
conventional SUSY signals would be especially interesting since it could
stringently restrict the underlying framework.
%

\section{Summary and Concluding Remarks}

The search for SUSY has become a standard item on all high energy
physics experiments searching for physics beyond the Standard
Model. For the most part, the analyses of current experiments as well as
projections of capabilities of future experiments have have been
carried out within the framework of the mSUGRA model, or within the MSSM
framework with some {\it ad hoc} assumptions motivated by mSUGRA 
about scalar and gaugino masses. Since SUSY cross sections, after experimental
cuts,
are expected to
be sensitive to sparticle mass and mixing patterns
(which are determined by the presently unkown dynamics of SUSY
breaking), it is worthwhile to examine just how much
the SUSY reach of future facilities change in alternative scenarios.

These considerations motivated us to examine SUSY signals at Tevatron
upgrades in the supergravity $SU(5)$ model \cite{anderson} with
non-universal gaugino masses at the GUT scale. For simplicity, other
parameters were considered to unify as in the mSUGRA model.  
Such a scenario can be realized if there is a superfield ${\Phi}$
that is charged under $SU(5)$ and whose auxiliary component develops a vev
that breaks the GUT gauge
group down to the Standard Model gauge group. The resulting GUT scale
gaugino masses are determined by the transformation properties of 
${\Phi}$, which can
transform as the {\bf 1} (this
corresponds to mSUGRA), {\bf 24}, {\bf 75} or the {\bf 200}
dimensional representation of
$SU(5)$. The resulting gaugino mass ratios at the GUT scale along with
their renormalized values at the weak scale (relevant for phenomenology)
are shown in
Table I. The phenomenology is altered not only because of the differences
in these weak scale gaugino masses, but also because the difference in the
boundary condition on gaugino masses alters the renormalization group
evolution of other parameters as well.

Our main result is the reach of Tevatron Main Injector and its possible
TeV33 luminosity upgrade for the cases where ${\Phi}$ belongs to any one
of these irreducible representations. We have examined this reach for
various event topologies. The results of our calculation are shown in
Figs.~\ref{FIG5}-\ref{FIG18}. The cumulative reach for SUSY, {\it i.e.}
the region of the $m_0- M_3^0$ plane 
where there should be an observable signal in at least one of
the channels, is shown in Figs \ref{FIG19} and \ref{FIG20}. 
The precise reach is model-dependent. In
Fig. \ref{FIG19}{\it a}, the reach in the mSUGRA model is built
entirely out of the reach in the J0L and C3L channels. For some of the
points examined, there may be observable signals in other channels
as well. The reach of the Tevatron for the
$F_{\Phi}\sim{\bf 24}$ model is built out of the J0L, C3L and JZ
channels, {\it i.e.} for a few points the SUSY signal appears to be
observable only via the JZ channel, and not in the more standard J0L and
C3L channels.  In addition, over much of the observable parameter space,
signals should also be detectable in many different leptonic
channels. The additional signals should help in constraining the
underlying model.
In contrast, the reach in the {\bf 75} and {\bf 200} models
shown in Fig. \ref{FIG19}{\it c}) and {\it d}), the cumulative reach
plot coincides with the reach plot for the J0L channel
(Fig. {\ref{FIG5})! In fact, the leptonic signals for SUSY will be
observable for only extremely restricted regions of model parameters.
This underscores the importance of the J0L channel in that it
is relatively model independent, at least so long as the LSP is a stable
neutralino which escapes detection: experimentalists should scrutinize this
channel closely even if no leptonic SUSY signals can be seen. We should
also mention that in our analysis, we have not attempted to really
optimize the reach in this channel. By judiciously choosing the cuts, it
may be possible to increase the reach somewhat beyond what appears in
the figure.

For the $\tan\beta =25$ case shown in Fig. \ref{FIG20}, the cumulative 
reach for
mSUGRA shown in frame {\it a}) is again defined by just the J0L and C3L
channels but is somewhat reduced relative to the corresponding
low $\tan\beta$ 
case. The reach for the $F_{\Phi}\sim{\bf 24}$ model shown in 
frame {\it b}) is again defined by the J0L, C3L and JZ channels, which
underscores the importance of an independent search for SUSY in the JZ channel.
The reach is only slightly diminished from the $\tan\beta =5$ case.
The Tevatron SUSY reach for the $F_{\Phi}\sim{\bf 75}$ and {\bf 200} models
for $\tan\beta =25$ is again defined solely by the J0L channel; very
few of these parameter space points are accessible in any other channel.
Thus, a SUSY discovery with a signal only in the J0L channel may indicate
non-universal gaugino masses which act to suppress leptonic signals
originating from SUSY particle cascade decays.

In summary, we have examined the SUSY reach of luminosity upgrades of
the Tevatron in non-minimal SUGRA type models where gaugino masses are
not unified at some high scale.  We find that rates for various signal
topologies (and hence, the reach) can be quite different from mSUGRA
expectations. There may be new signatures such as the high $p_T$ $Z +
\eslt$ signal in the {\bf 24} model that are unobservable in the mSUGRA
picture. On the other hand, in the {\bf 75} and {\bf 200} models a
signal might be observable only in the canonical multijet + $\eslt$
channel. This is in contrast to $R$-parity violating models \cite{RPV}
where there might be observable signals {\it only} in the multilepton
channel, but no signal in the usual $\eslt$ channel. We thus conclude that
while it might well be possible to discover a signal for new physics at
the Tevatron, its interpretation will have to be done with care. What we
do not see, in addition of course to what we do see, may play an
important role in unravelling the nature of the new physics.

%
\acknowledgments
This research was supported in part by the U.~S. Department of Energy
under contract number DE-FG02-97ER41022, DE-FG-03-94ER40833 
and DE-FG02-91-ER4086. 

%
%

\newpage
%
%

\iftightenlines\else\newpage\fi
\iftightenlines\global\firstfigfalse\fi
\def\dofig#1#2{\epsfxsize=#1\centerline{\epsfbox{#2}}}

\begin{table}
\begin{center}
\begin{small}
\begin{tabular}{|c|ccc|ccc|}
\hline
\ & \multicolumn{3}{c|} {$\mgut$} & \multicolumn{3}{c|}{$\mz$} \cr
$F_{\Phi}$

& $M_3$ & $M_2$ & $M_1$

& $M_3$ & $M_2$ & $M_1$ \cr
\hline

${\bf 1}$   & $1$ &$\;\; 1$  &$\;\;1$   & $\sim \;6$ & $\sim \;\;2$ &

$\sim \;\;1$ \cr
${\bf 24}$  & $2$ &$-3$      & $-1$  & $\sim 12$ & $\sim -6$ &

$\sim -1$ \cr
 ${\bf 75}$  & $1$ & $\;\;3$  &$-5$      & $\sim \;6$ & $\sim \;\;6$ &

$\sim -5$ \cr
${\bf 200}$ & $1$ & $\;\; 2$ & $\;10$   & $\sim \;6$ & $\sim \;\;4$ &

$\sim \;10$ \cr
\hline
\end{tabular}
\end{small}
\smallskip
\caption{Relative gaugino masses at $\mgut$ and $\mz$
in  the four possible $F_{\Phi}$ irreducible representations.}
\label{masses}
\end{center}
\end{table}
\newpage
%


\begin{figure}
\dofig{6in}{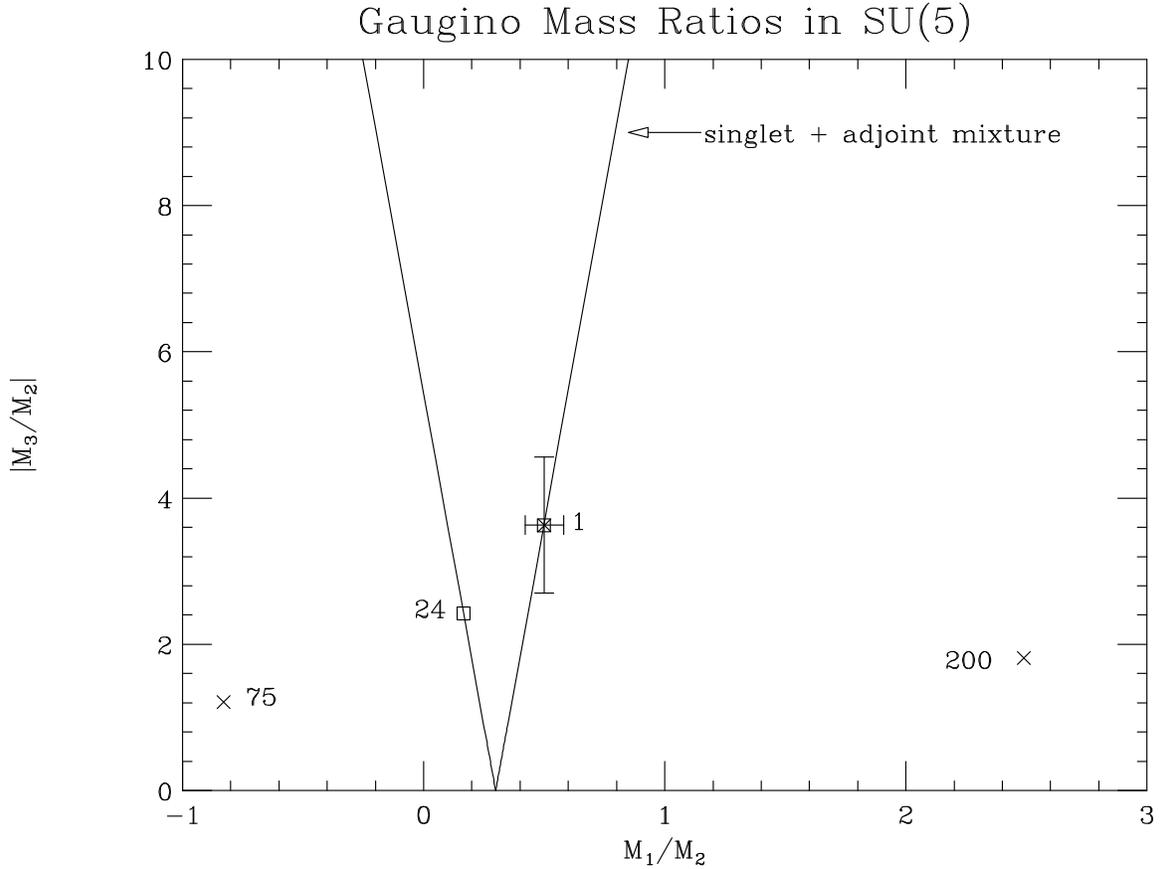}
\caption[]{
Gaugino mass ratios for the four special cases where the field
$\Phi$ transforms as the {\bf 1}, {\bf 24}, {\bf 75} or {\bf 200}
dimensional representation of $SU(5)$, or as
an arbitrary linear combination of the singlet and adjoint
representations (solid line).
}
\label{fig0}
\end{figure}

\begin{figure}
\dofig{6in}{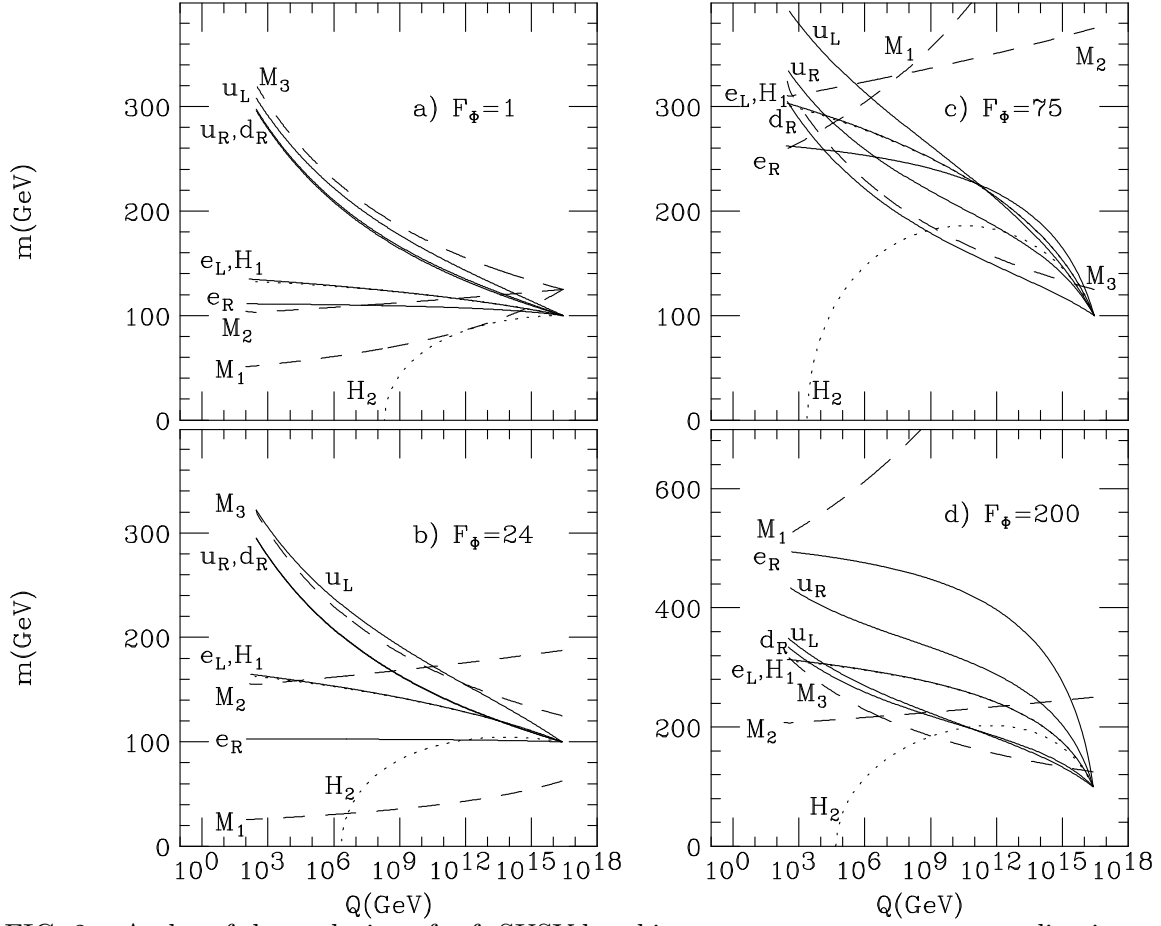}
\caption[]{
A plot of the evolution of soft SUSY breaking parameters versus
renormalization scale $Q$ from $M_{GUT}$ to $M_{weak}$ for SUGRA model
parameters $m_0=100$ GeV, $M_3^0=125$ GeV, $A_0=0$, $\tan\beta =5$
and $\mu >0$, for the {\it a}) $F_\Phi \sim 1$, 
{\it b}) $F_\Phi\sim 24$, {\it c}) $F_\Phi\sim 75$ and 
{\it d}) $F_\Phi\sim 200$ models.
We take $m_t=175$ GeV. Notice the different scale in frame {\it d})}
\label{FIG1}
\end{figure}
\begin{figure}
\dofig{6in}{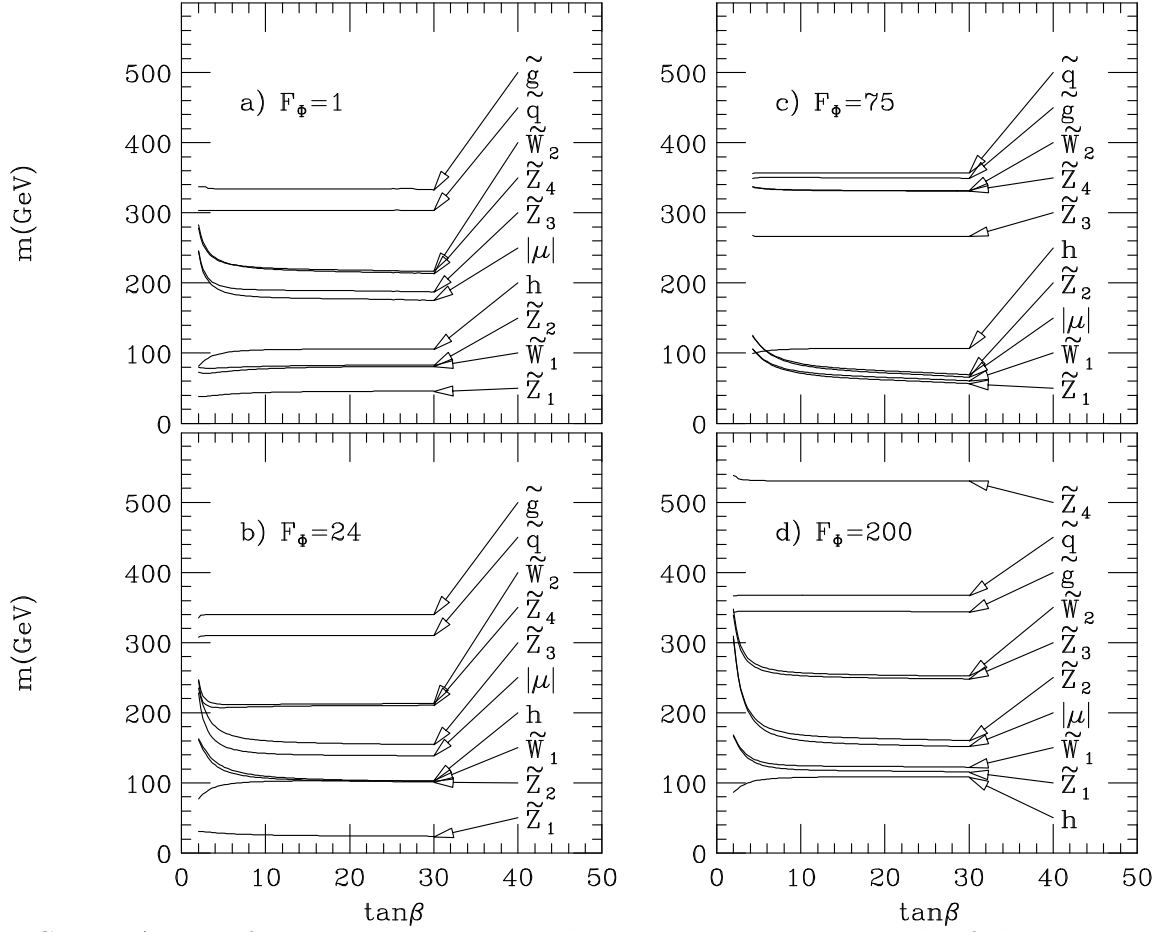}
\caption[]{
A plot of various physical sparticle masses and the magnitude of the $\mu$
parameter versus $\tan\beta$ for
SUGRA model parameters $m_0=100$ GeV, $M_3^0=125$ GeV, $A_0=0$
and $\mu >0$, for the {\it a}) $F_\Phi\sim 1$, 
{\it b}) $F_\Phi\sim 24$, {\it c}) $F_\Phi\sim 75$ and 
{\it d}) $F_\Phi\sim 200$ models.
The squark mass is averaged over the first two generations.}
\label{FIG2}
\end{figure}
\begin{figure}
\dofig{6in}{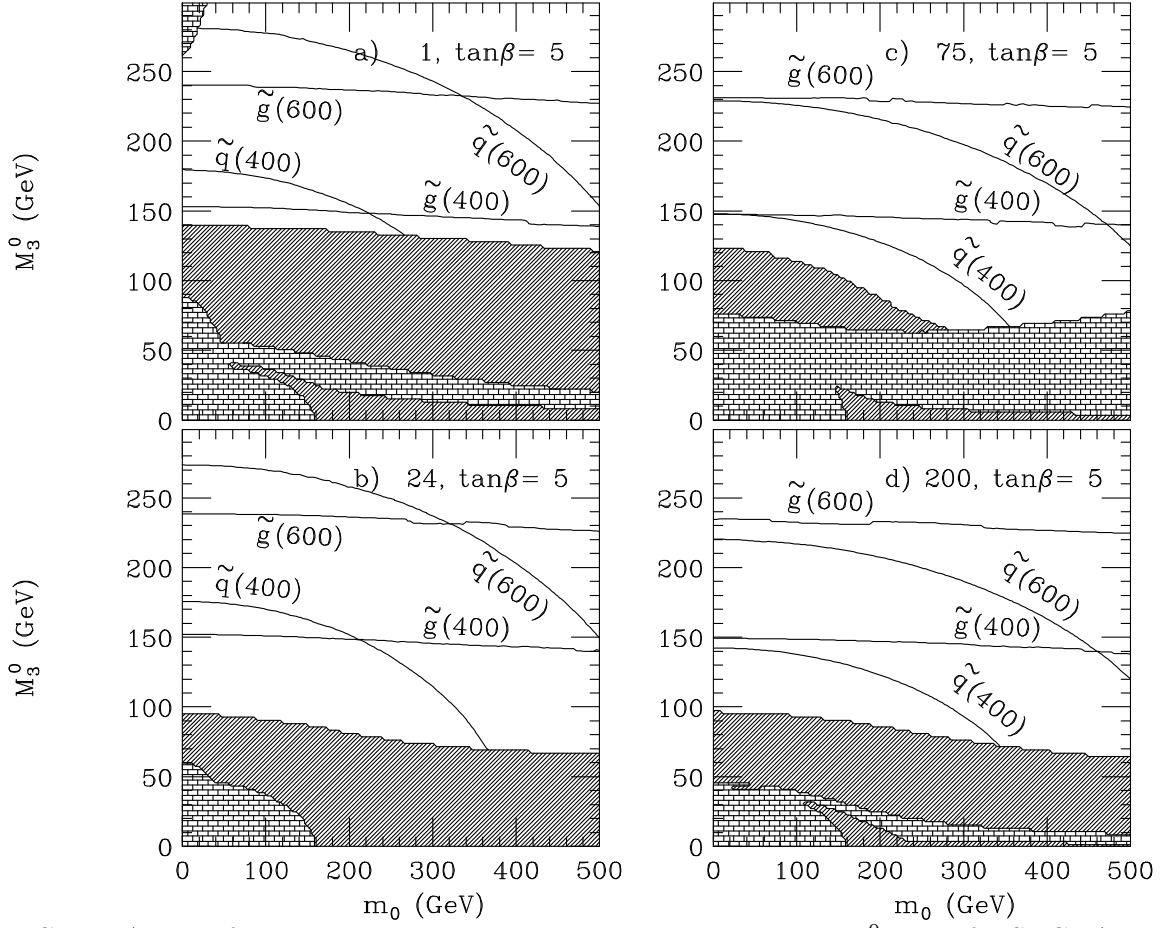}
\caption[]{
A plot of gluino and squark mass contours in the 
$m_0\ vs.\ M_3^0$ plane for 
SUGRA model parameters $A_0=0$, $\tan\beta =5$ and $\mu >0$ 
for the {\it a}) $F_\Phi \sim 1$, 
{\it b}) $F_\Phi\sim 24$, {\it c}) $F_\Phi\sim 75$ and 
{\it d}) $F_\Phi\sim 200$ models.
The squark mass is averaged over the first generation.}
\label{FIG3}
\end{figure}
\begin{figure}
\dofig{6in}{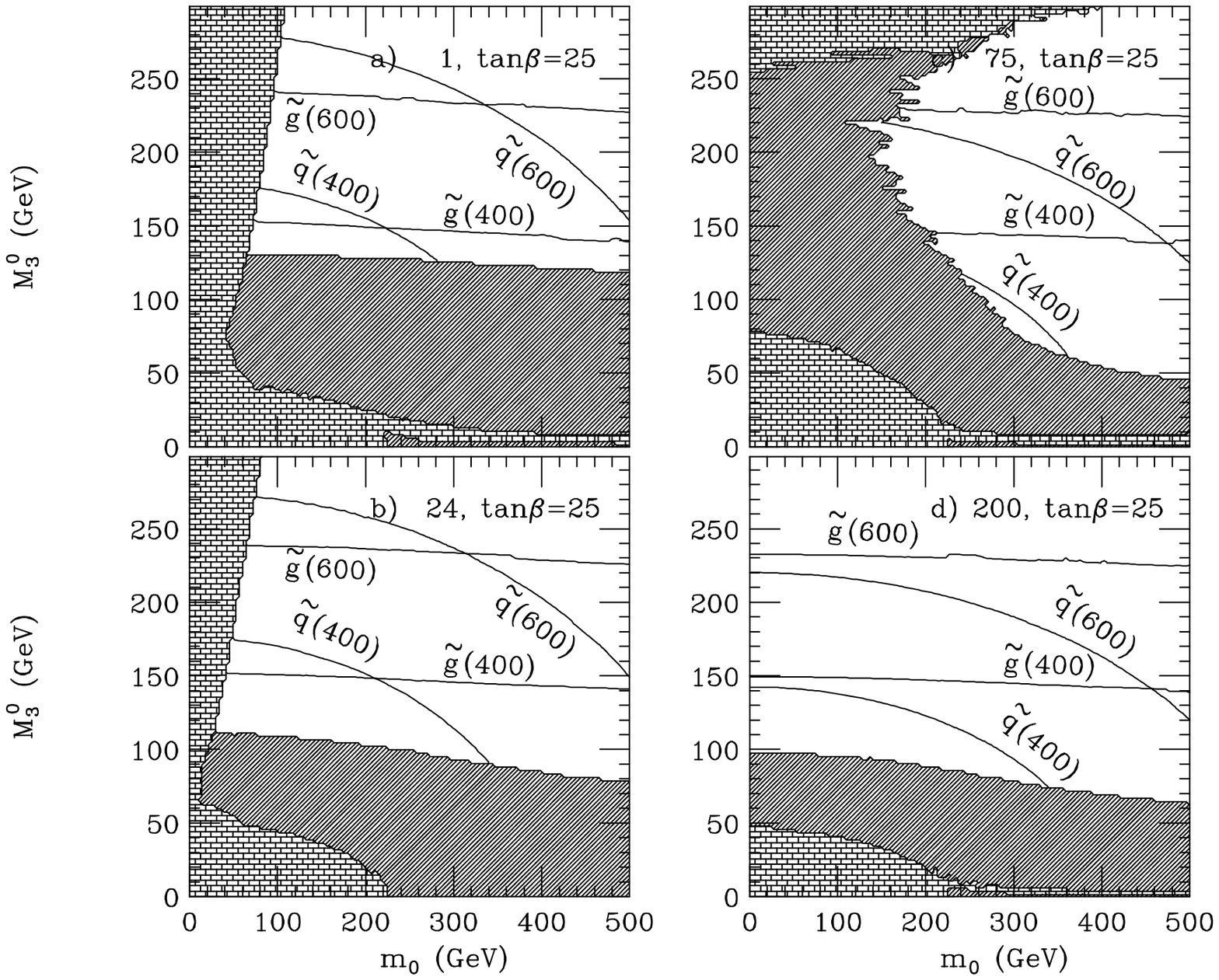}
\caption[]{
Same as Fig. \ref{FIG3} except for $\tan\beta =25$.}
\label{FIG4}
\end{figure}
\begin{figure}
\dofig{6in}{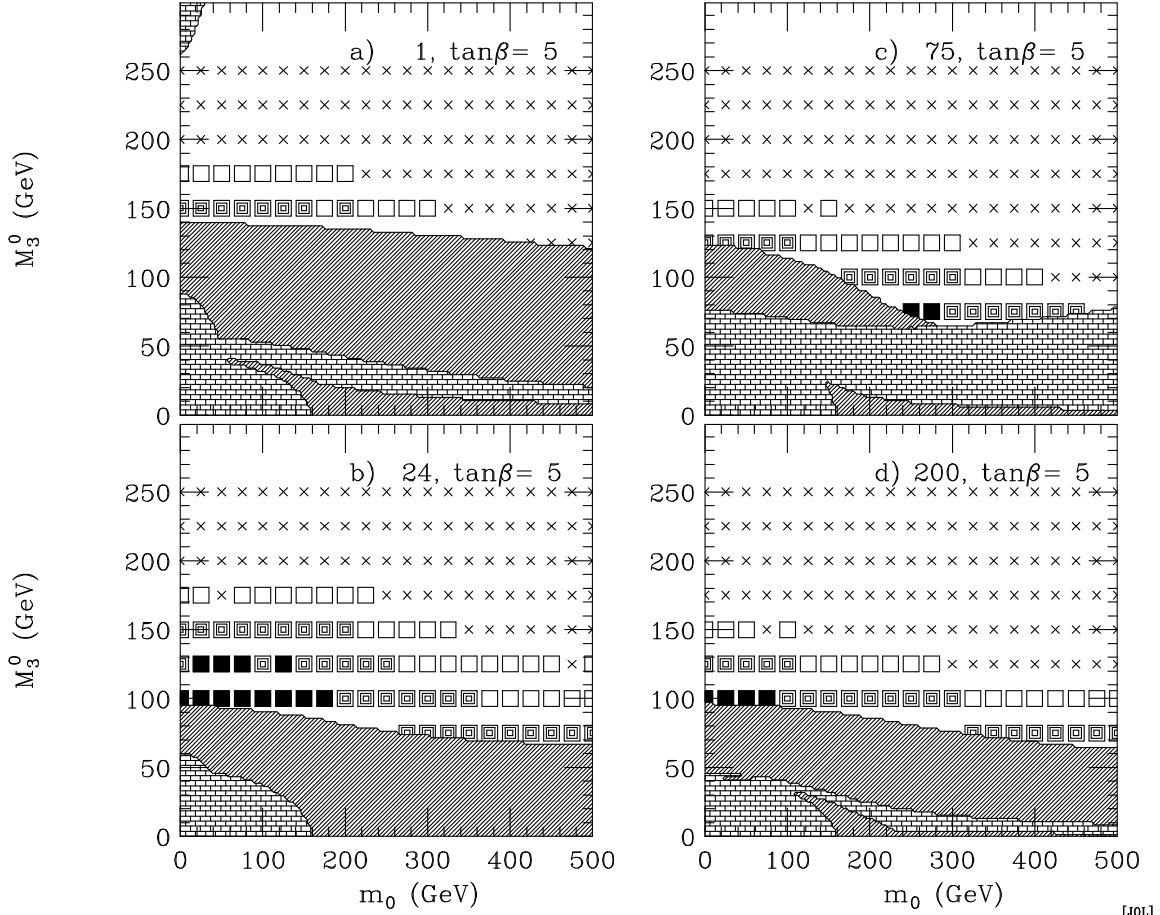}
\caption[]{
A plot of parameter space points accessible to Fermilab Tevatron
collider experiments with integrated luminosity 0.1 fb$^{-1}$ (black squares),
2 fb$^{-1}$ (gray squares) and 25 fb$^{-1}$ (white squares) via
the multijet $+\eslt$ signal (J0L). Events containing isolated leptons
have been vetoed.
Points are plotted in the
$m_0\ vs.\ M_3^0$ plane for 
SUGRA model parameters $A_0=0$, $\tan\beta =5$ and $\mu >0$ 
for the {\it a}) $F_\Phi \sim 1$, 
{\it b}) $F_\Phi\sim 24$, {\it c}) $F_\Phi\sim 75$ and 
{\it d}) $F_\Phi\sim 200$ models.}
\label{FIG5}
\end{figure}
\begin{figure}
\dofig{6in}{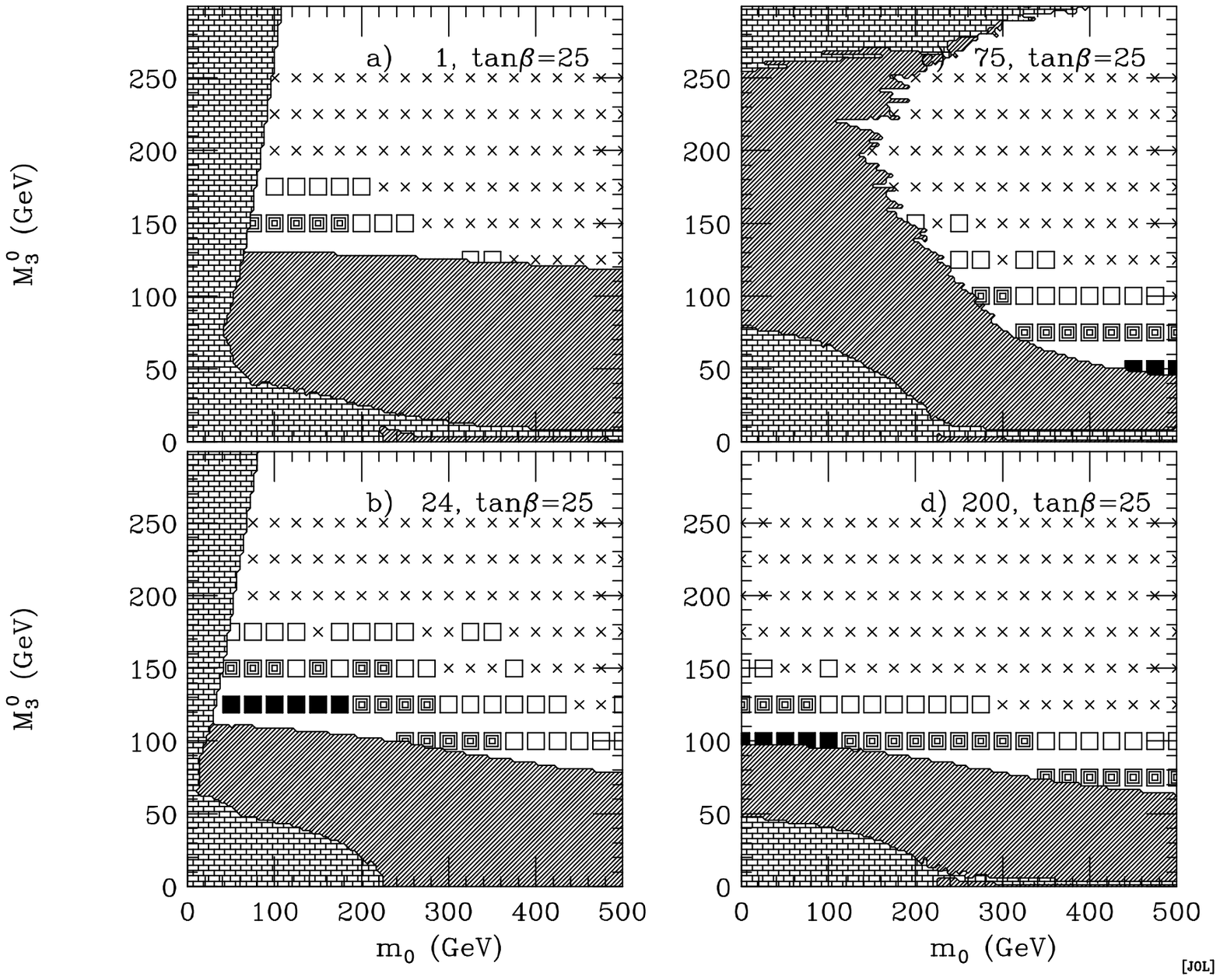}
\caption[]{
The same as Fig. \ref{FIG5}, except for $\tan\beta =25$.}
\label{FIG6}
\end{figure}
\begin{figure}
\dofig{6in}{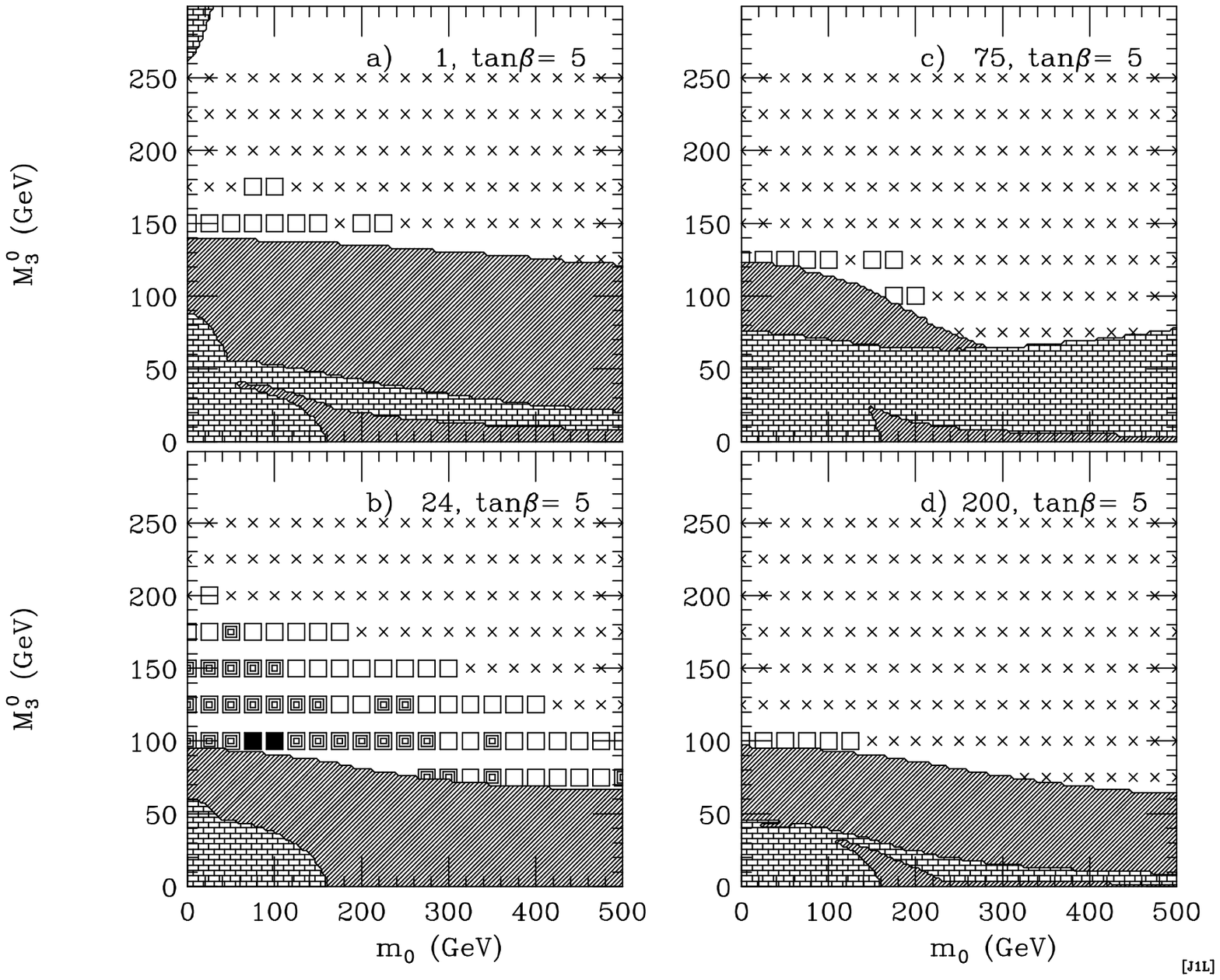}
\caption[]{
The same as Fig. \ref{FIG5}, except for the J1L signal.}
\label{FIG7}
\end{figure}
\begin{figure}
\dofig{6in}{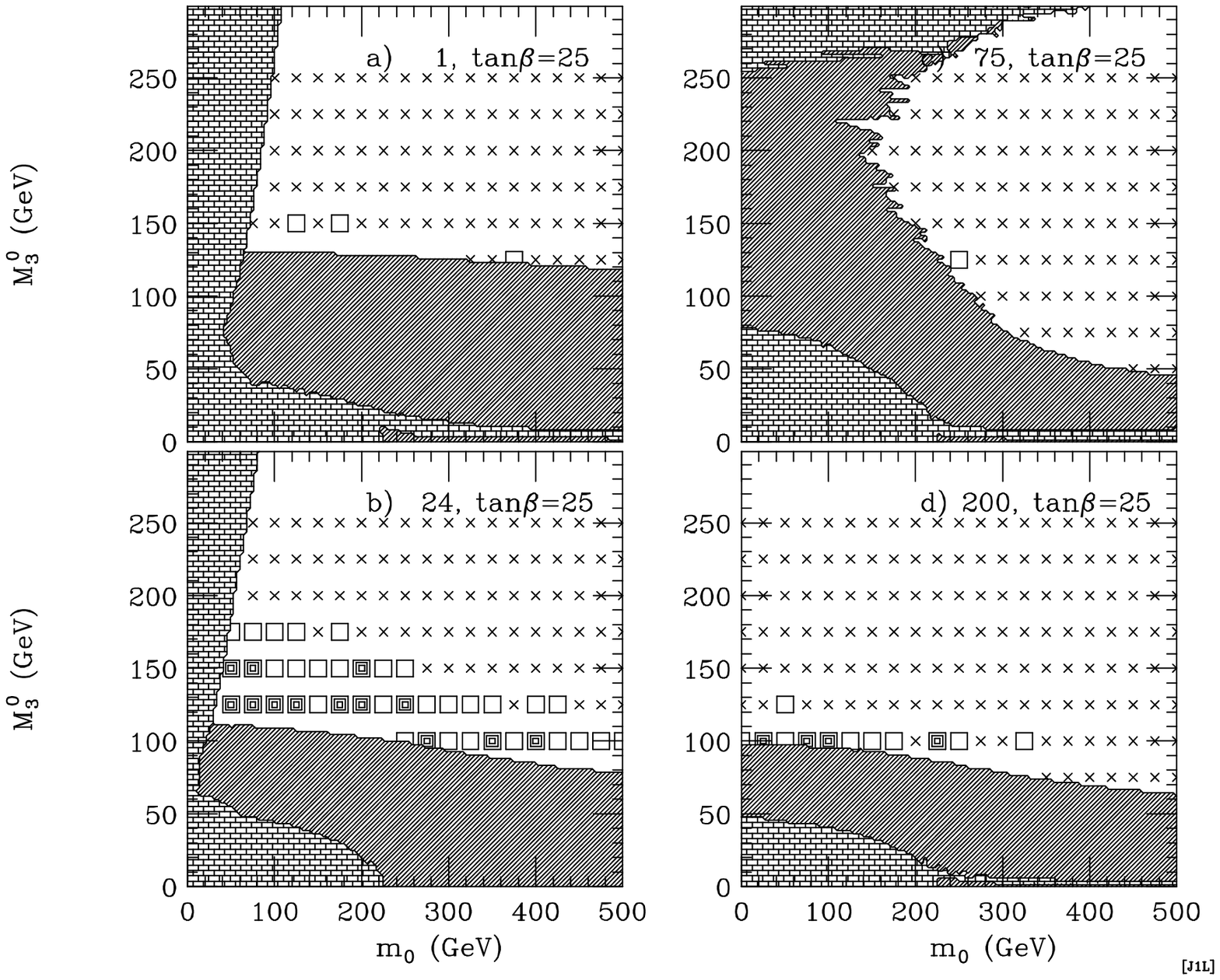}
\caption[]{
The same as Fig. \ref{FIG6}, except for the J1L signal.}
\label{FIG8}
\end{figure}
\begin{figure}
\dofig{6in}{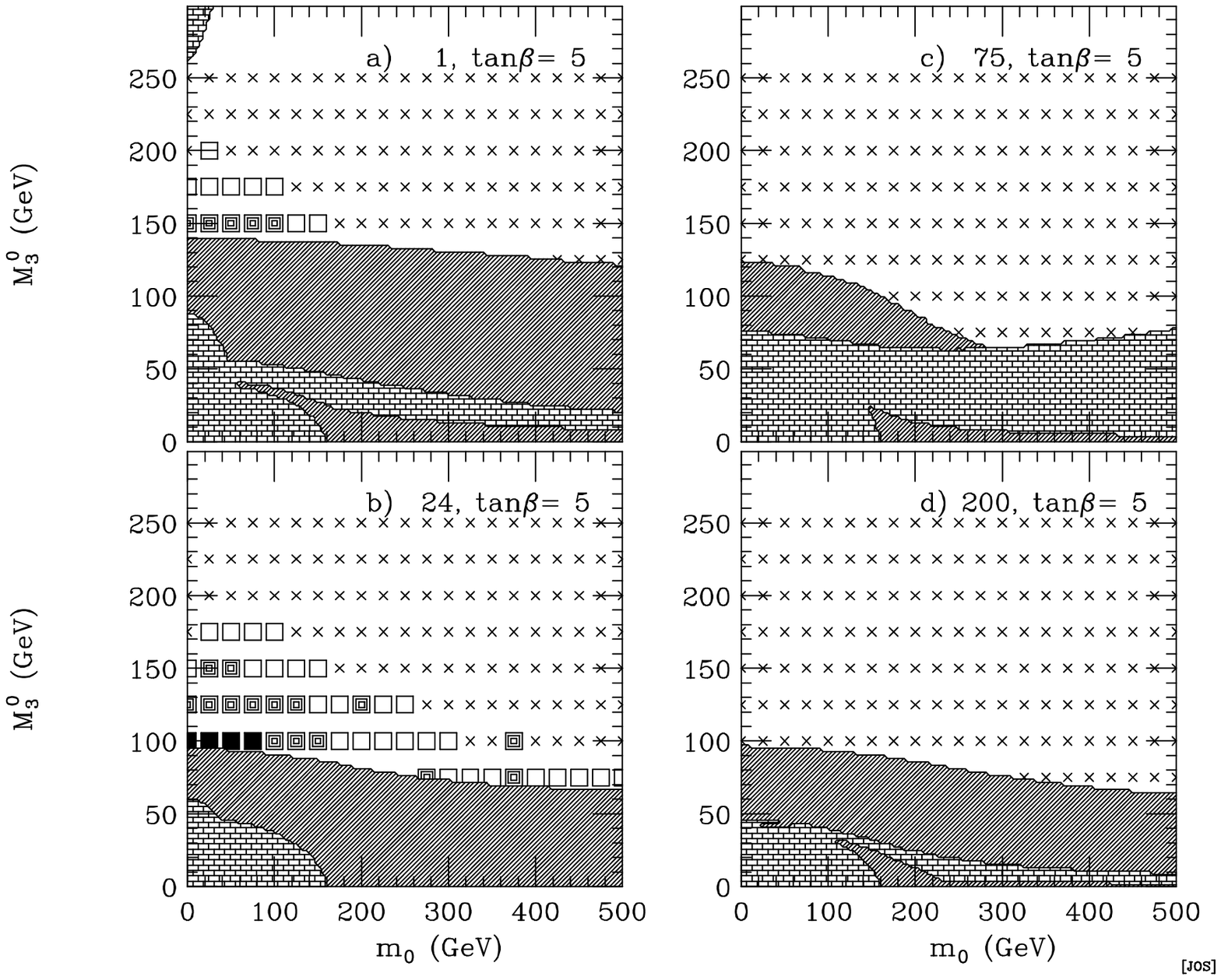}
\caption[]{
The same as Fig. \ref{FIG5}, except for the JOS signal.}
\label{FIG9}
\end{figure}
\begin{figure}
\dofig{6in}{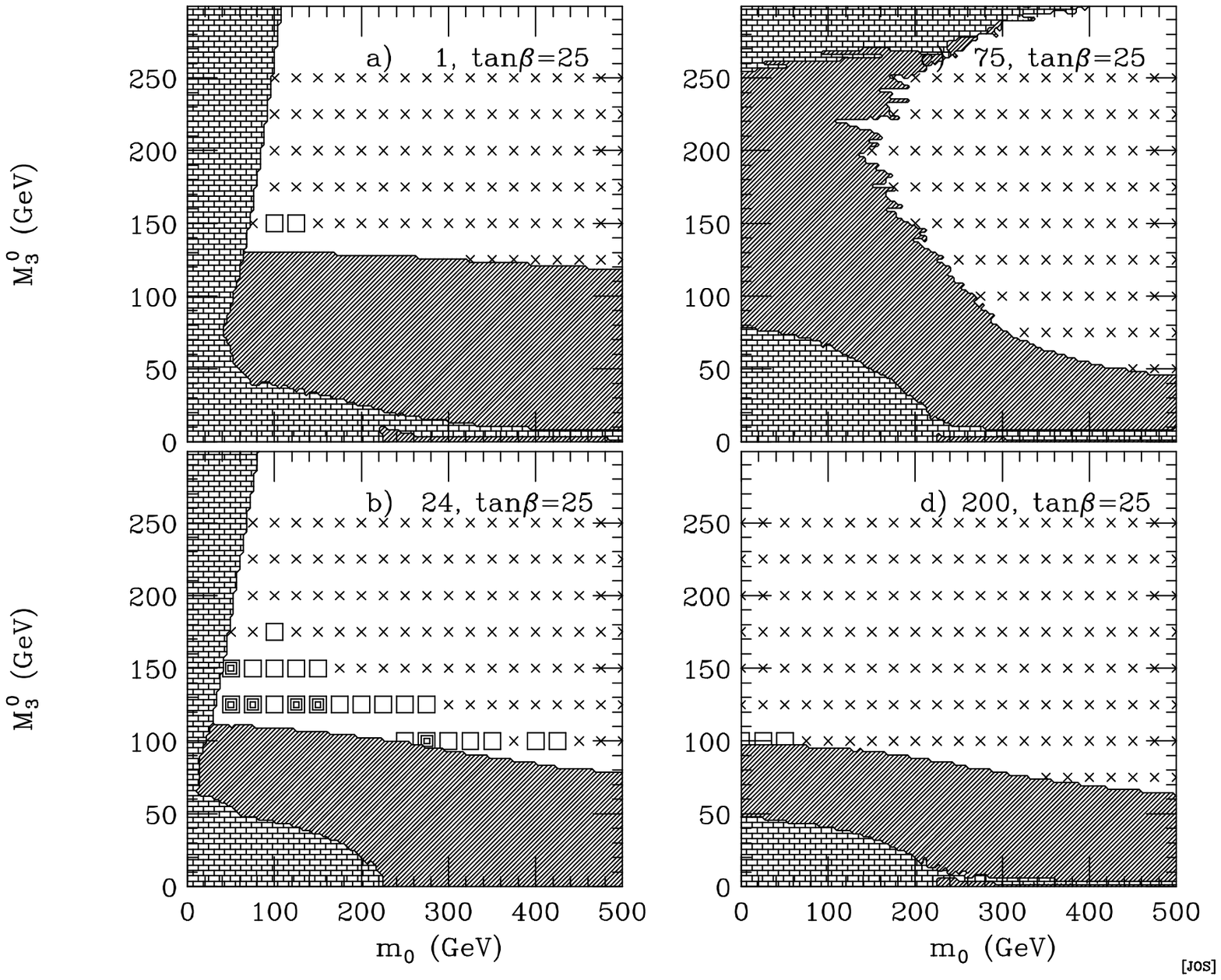}
\caption[]{
The same as Fig. \ref{FIG6}, except for the JOS signal.}
\label{FIG10}
\end{figure}
\begin{figure}
\dofig{6in}{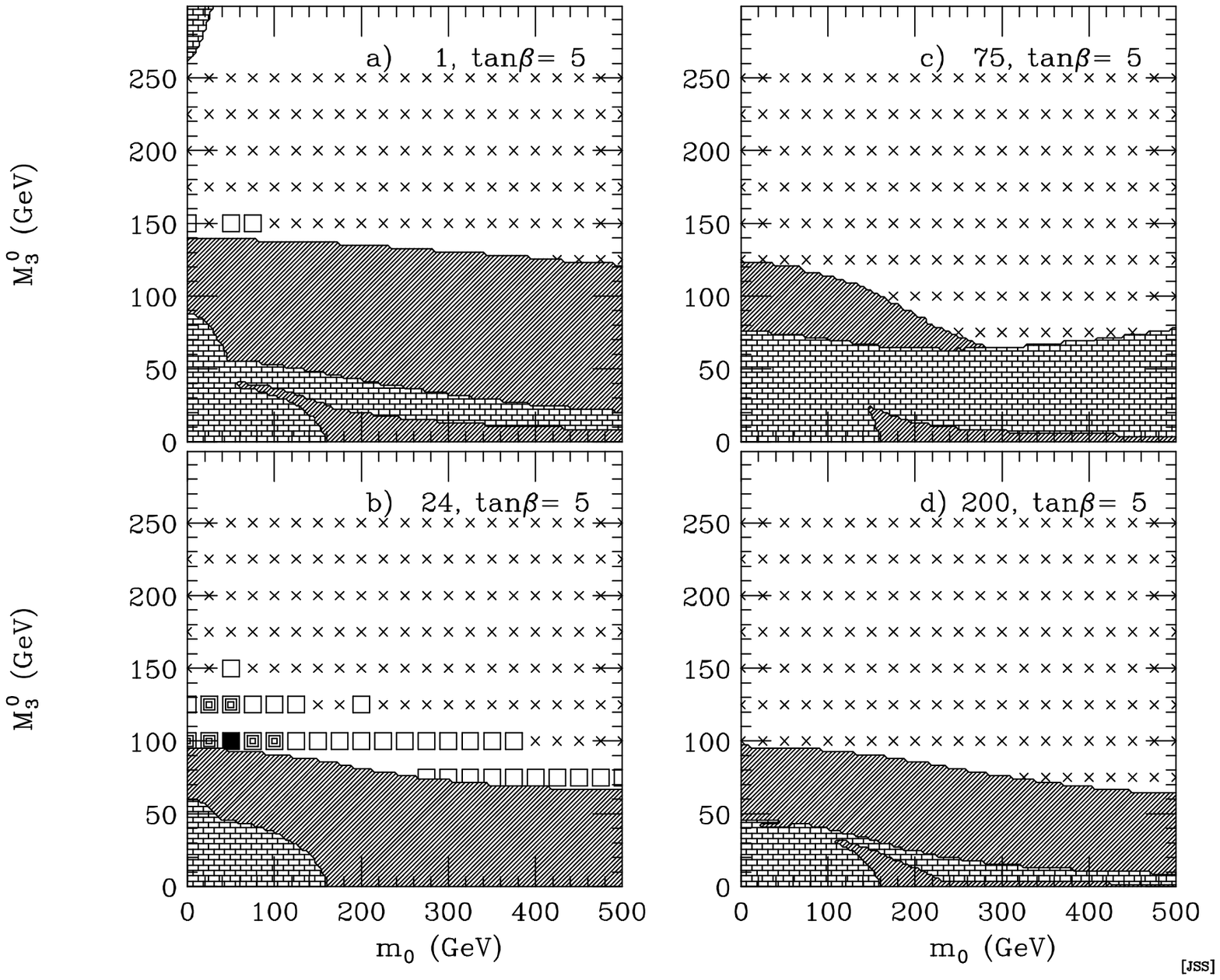}
\caption[]{
The same as Fig. \ref{FIG5}, except for the JSS signal.}
\label{FIG11}
\end{figure}
\begin{figure}
\dofig{6in}{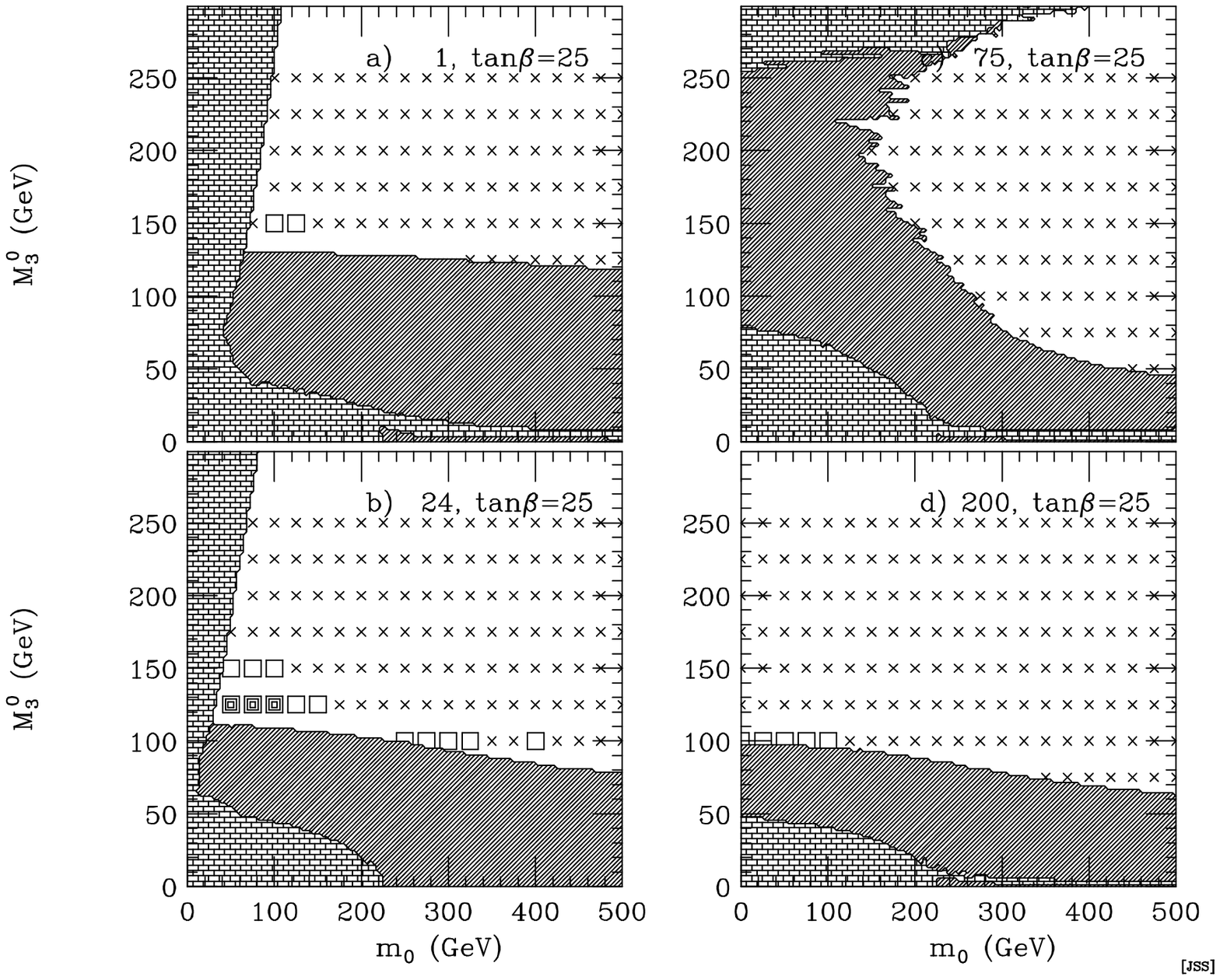}
\caption[]{
The same as Fig. \ref{FIG6}, except for the JSS signal.}
\label{FIG12}
\end{figure}
\begin{figure}
\dofig{6in}{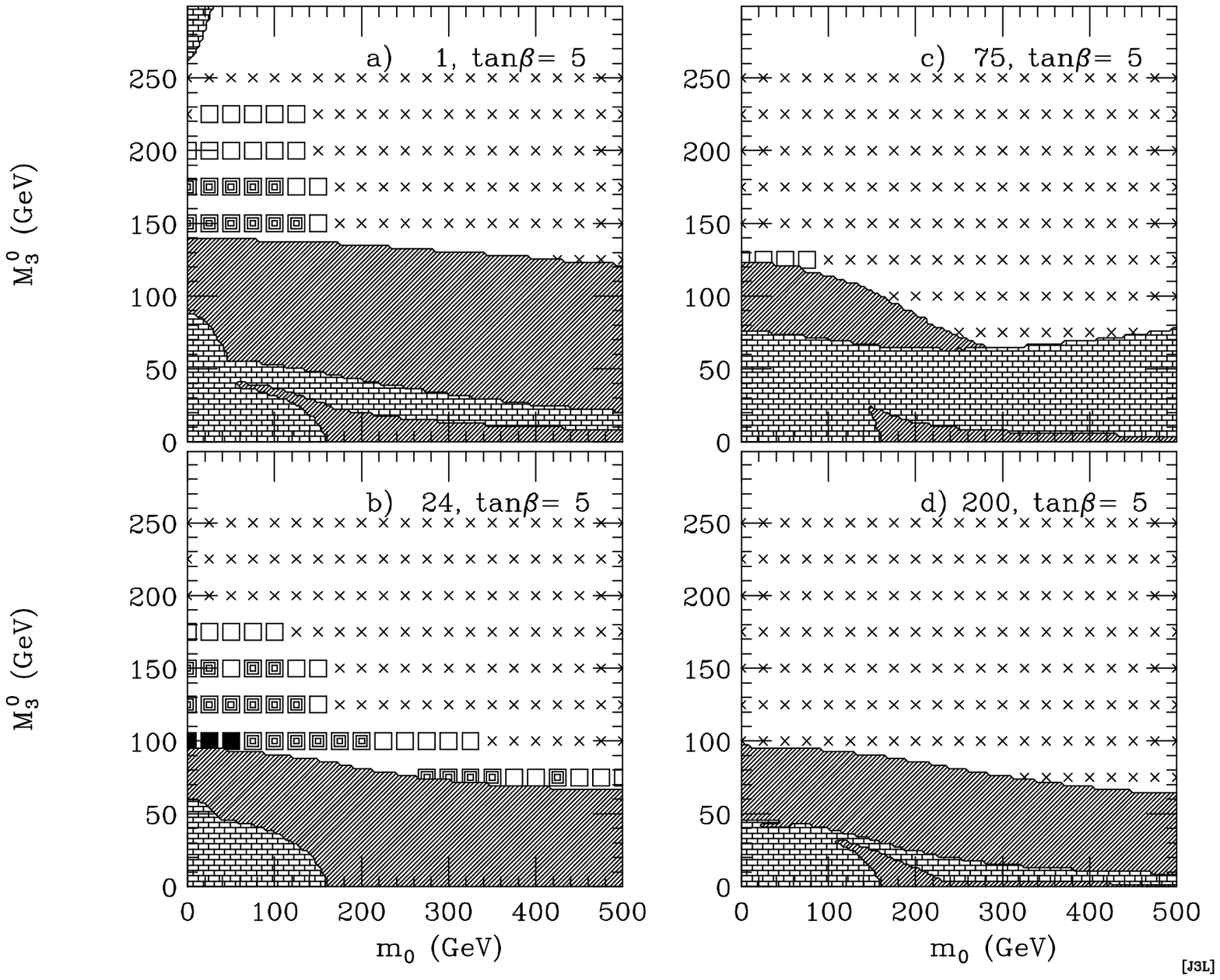}
\caption[]{
The same as Fig. \ref{FIG5}, except for the J3L signal.}
\label{FIG13}
\end{figure}
\begin{figure}
\dofig{6in}{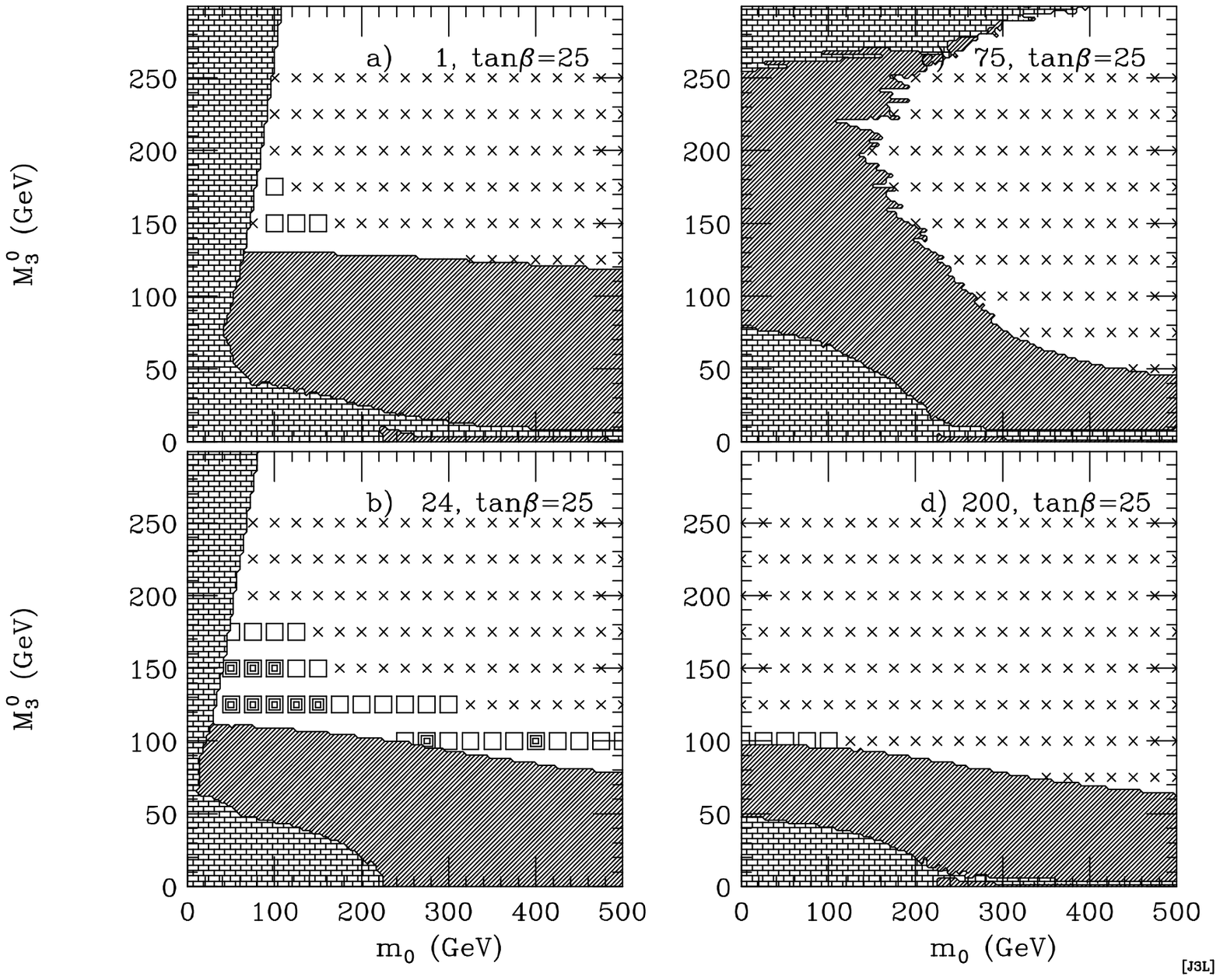}
\caption[]{
The same as Fig. \ref{FIG6}, except for the J3L signal.}
\label{FIG14}
\end{figure}
\begin{figure}
\dofig{6in}{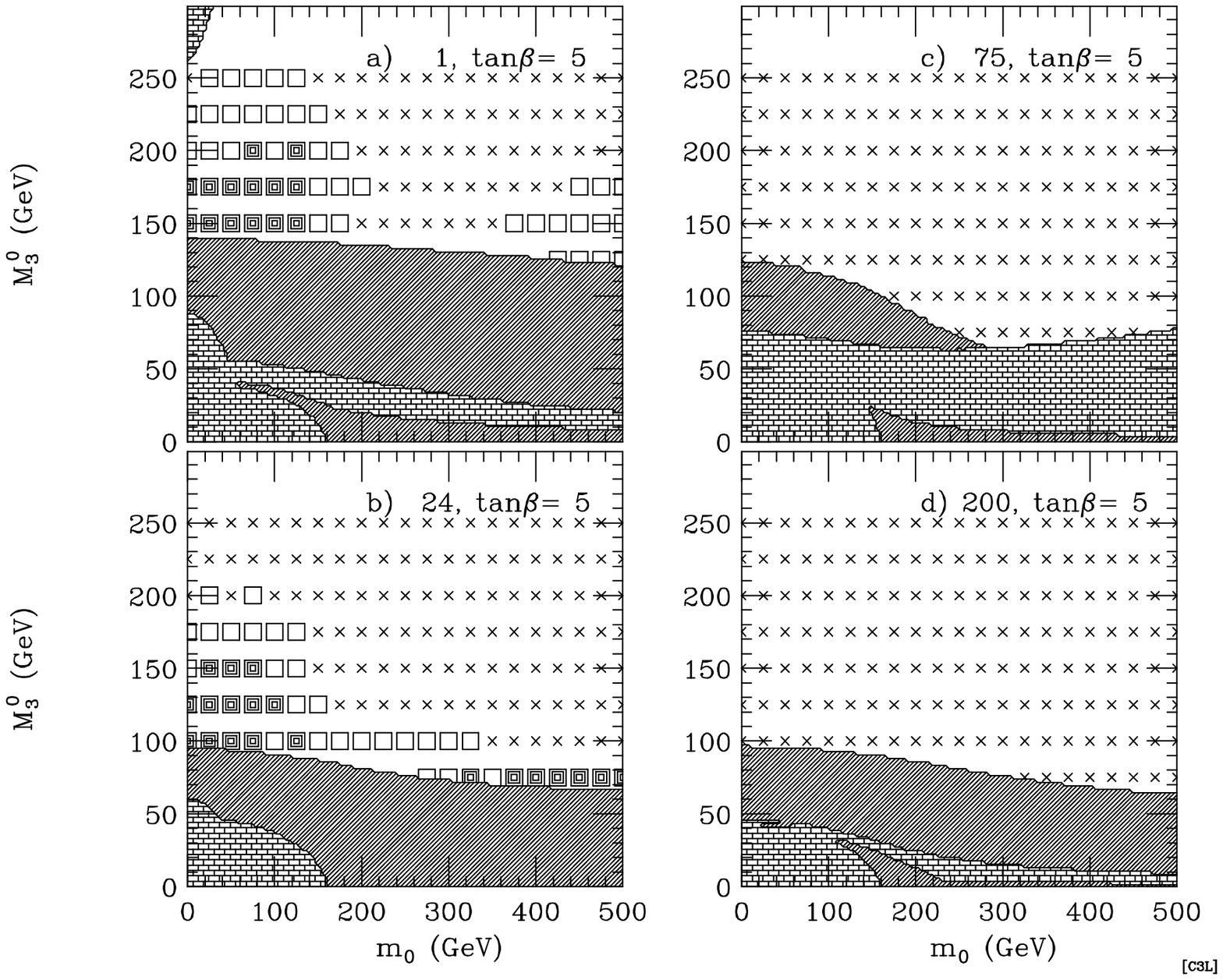}
\caption[]{
The same as Fig. \ref{FIG5}, except for the C3L signal.}
\label{FIG15}
\end{figure}
\begin{figure}
\dofig{6in}{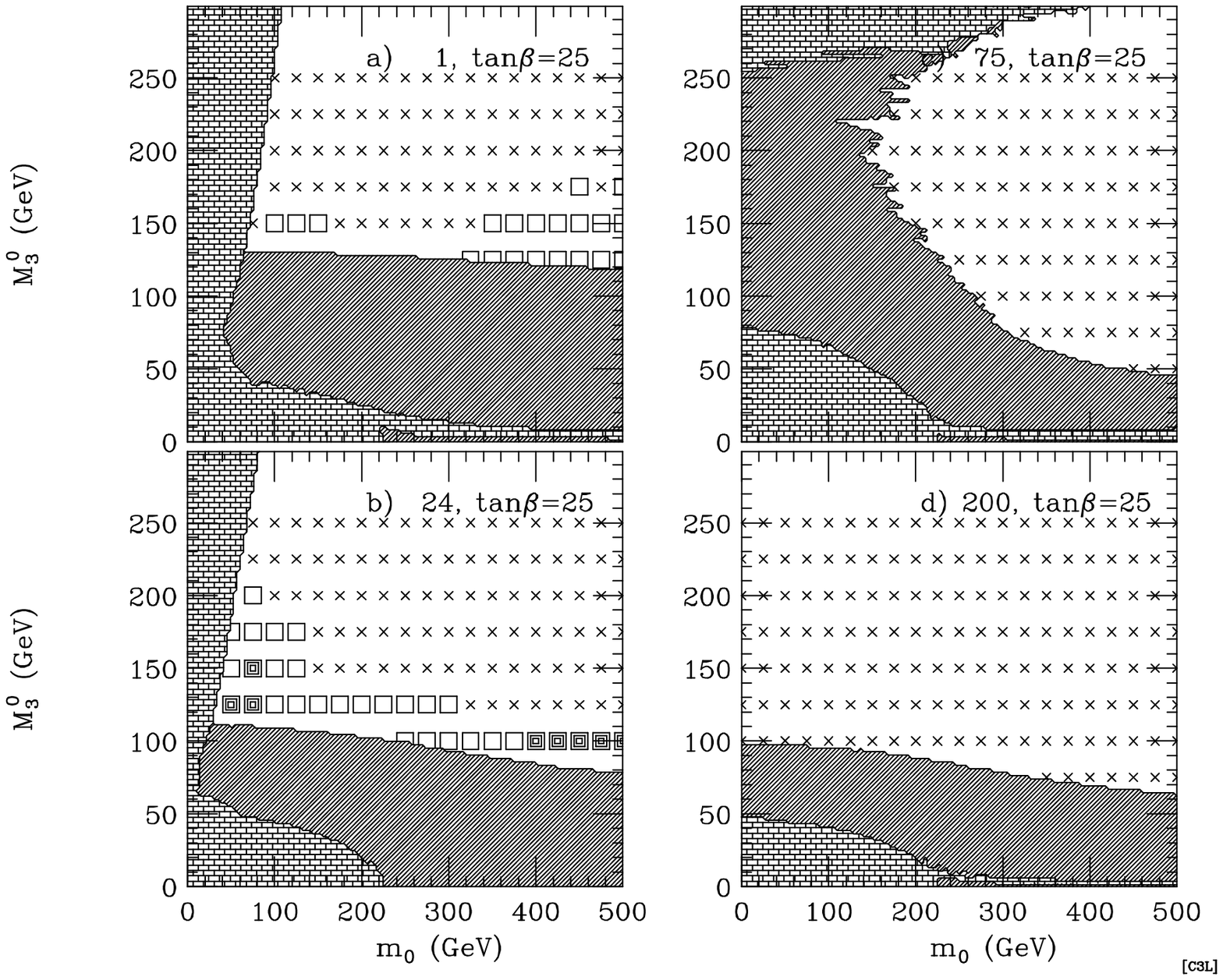}
\caption[]{
The same as Fig. \ref{FIG6}, except for the C3L signal.}
\label{FIG16}
\end{figure}
\begin{figure}
\dofig{6in}{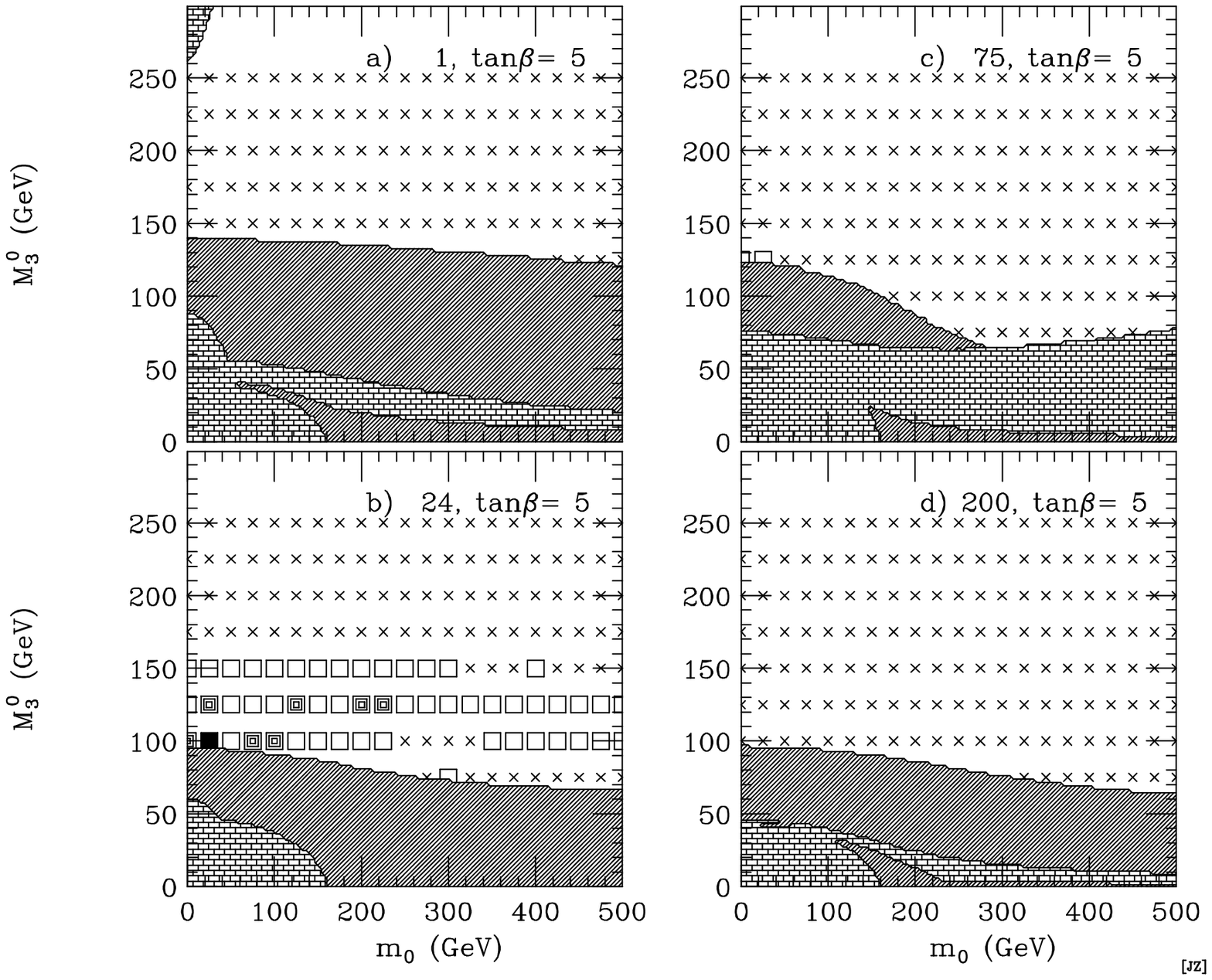}
\caption[]{
The same as Fig. \ref{FIG5}, except for the JZ signal.}
\label{FIG17}
\end{figure}
\begin{figure}
\dofig{6in}{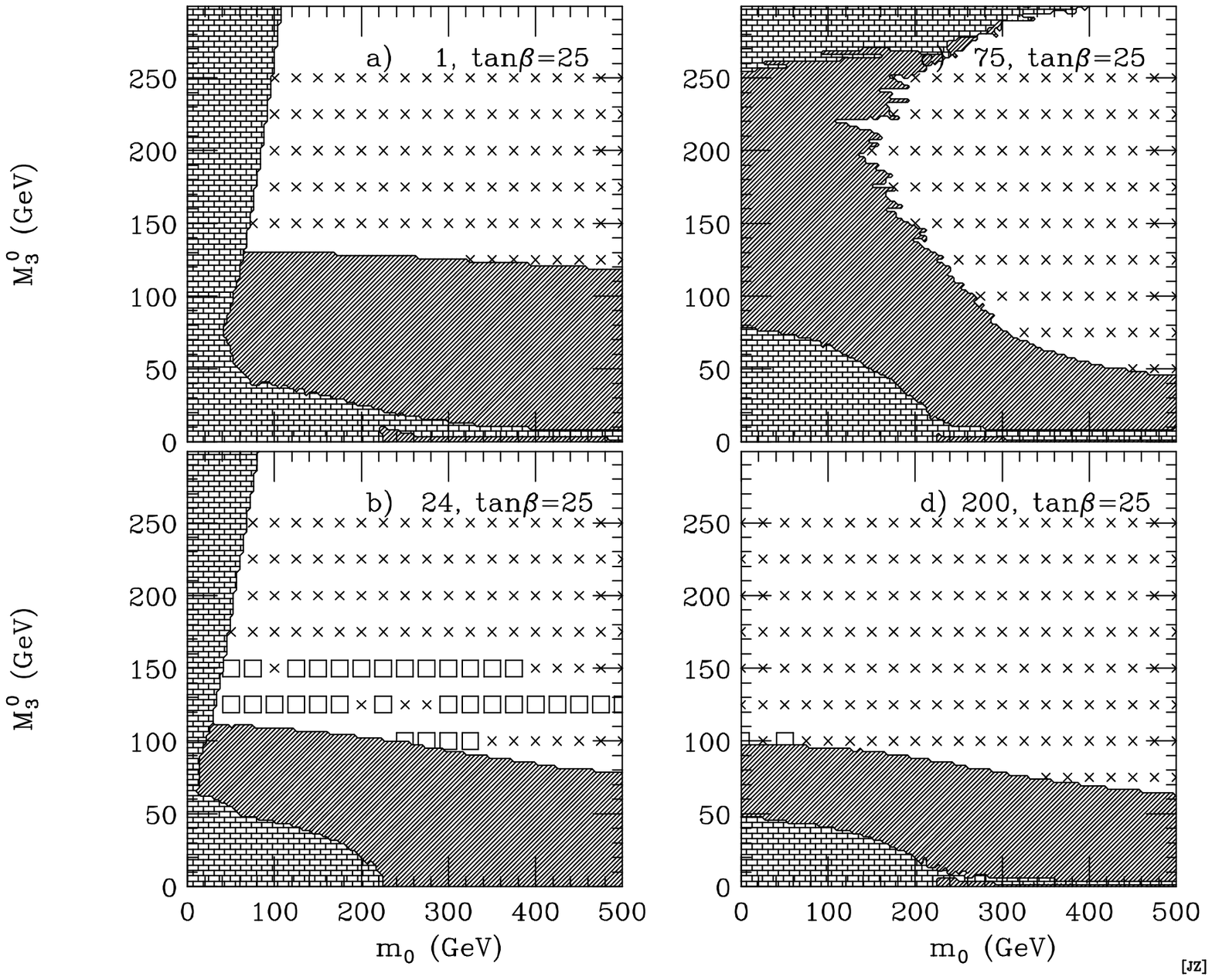}
\caption[]{
The same as Fig. \ref{FIG6}, except for the JZ signal.}
\label{FIG18}
\end{figure}
\begin{figure}
\dofig{6in}{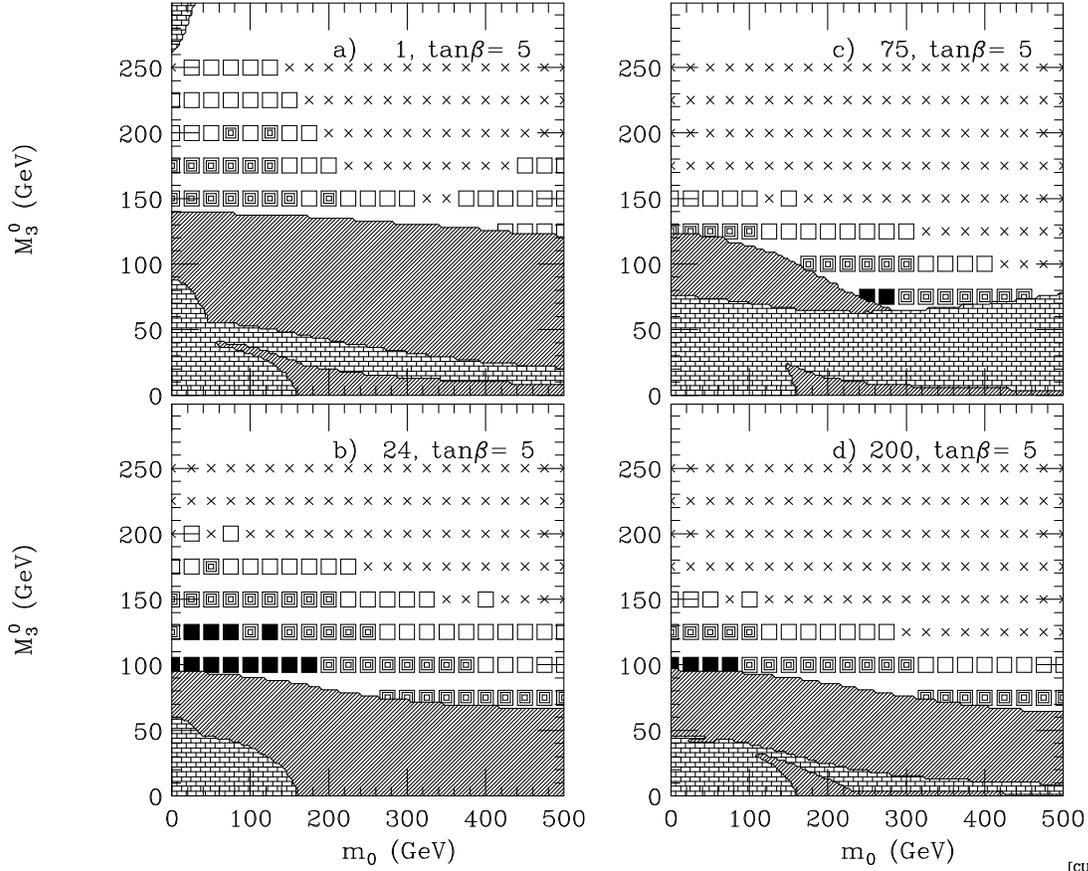}
\caption[]{
The same as Fig. \ref{FIG5}, except the reach is plotted 
for SUGRA models via {\it any} of the signals
considered in this paper.}
\label{FIG19}
\end{figure}
\begin{figure}
\dofig{6in}{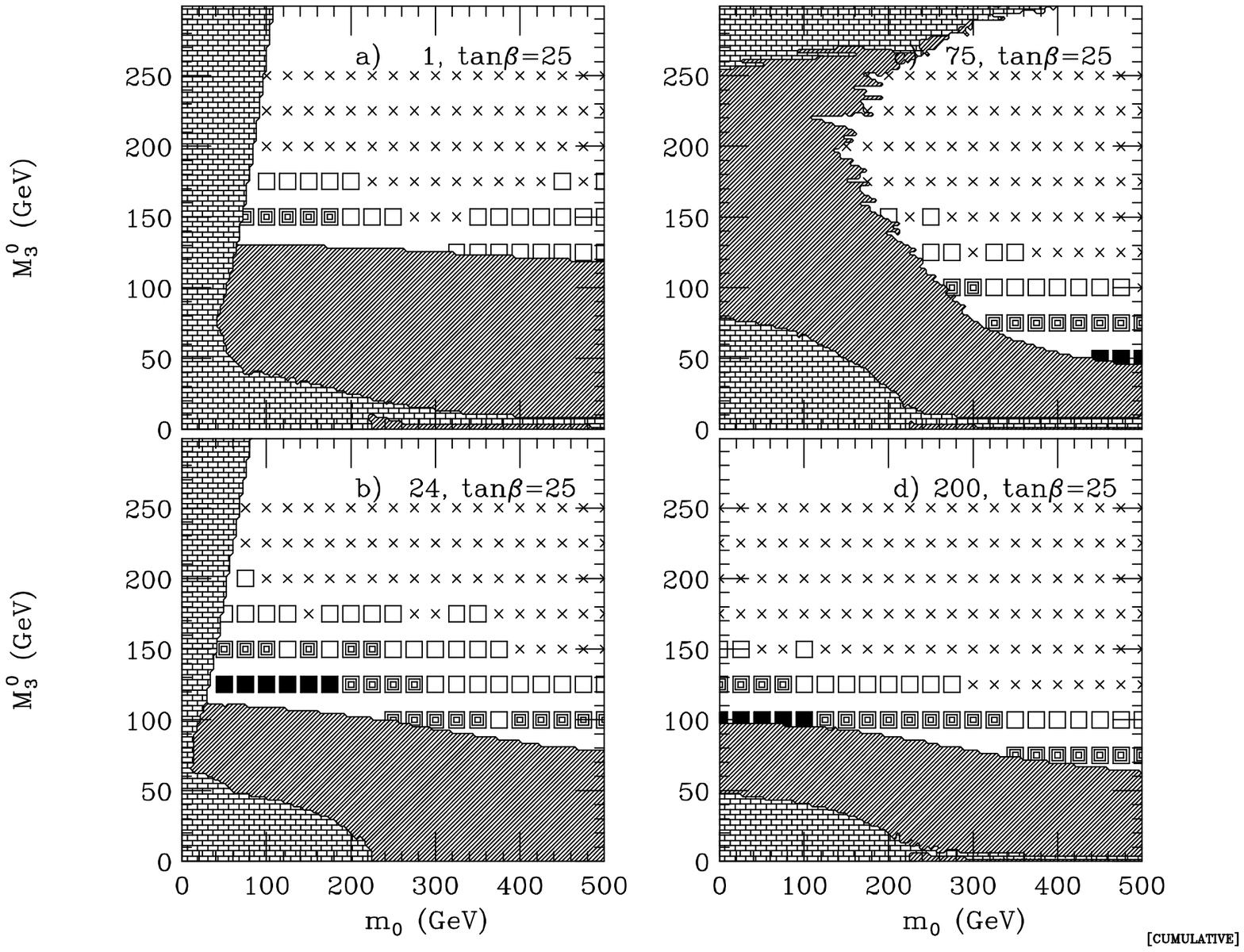}
\caption[]{
The same as Fig. \ref{FIG19}, except for $\tan\beta =25$.}
\label{FIG20}
\end{figure}

\end{document}